\documentclass[epsfig]{aa} 
\usepackage{graphics}
\newcommand{\convolution}{\mbox{$\ast$}}
\newcommand{\micron}{\mbox{$\mu m$~}}
\newcommand{\rc}{\mbox{$r_c$}}
\newcommand{\rt}{\mbox{$r_t$}}
\newcommand{\kms}{km\thinspace s$^{-1}$}
\newcommand{\q}[2]{\mbox{$Q_{#1}^{#2}$}}
\newcommand{\fig}[1]{Fig. \ref{#1}}
\newcommand \conc   {$c$ = log ($r_t/r_c$)}
\newcommand \msun   {\mbox{ M$_\odot$}}
\newcommand \etal   {et al.}
\newcommand \degree  {\mbox{$^\circ$}}
\newcommand{\tab}[1]{Table \ref{#1}}

\begin{document}

   \thesaurus{10     
              (03.20.1;  
	       09.04.1;  
	       10.07.2;  
               10.07.3;  
               10.19.3;  
               12.12.1)} 

\title{Tidal Tails Around 20 Galactic Globular Clusters: \thanks{Based
on observations made at  the European Southern Observatory, La  Silla,
Chile}\fnmsep \thanks{Plate   scanning  done  with  the MAMA  (Machine
Automatique \`a Mesurer pour l'Astronomie),  a facility developed  and
operated by the INSU (Institut National des  Sciences de l'Univers) at
the Observatoire de Paris, France}}

\subtitle{Observational    Evidence  for    Gravitational   Disk/Bulge
Shocking.}

\author{St\'ephane   Leon   \inst{1,2,3},  Georges  Meylan    \inst{4}  \&
Fran\c{c}oise Combes \inst{1}}
\offprints{S. Leon ({\em stephane.leon@obspm.fr})}

\institute{  DEMIRM, Observatoire de Paris, 
		61, Avenue de l'Observatoire,
		F-75014 Paris, France
		\and
		CAI,
		Observatoire de Paris,
		61, Avenue de l'Observatoire,
		F-75014 Paris, France
		\and
		Institute of Astronomy and Astrophysics, 
		Academia Sinica, P.O. Box 1-87, Nankang, 
		Taipei, Taiwan
		\and
		ESO, 
		Karl-Schwarzschild-Strasse 2, 
		D-85748 Garching, Germany
          }

\date{Received ; accepted }

\authorrunning{S. Leon, G. Meylan, \& F. Combes}

\titlerunning{Tidal Tails Around Globular Clusters}

\maketitle

\begin{abstract}
Large-field multi-color  images of 20  galactic globular  clusters are
used to investigate  the presence of tidal  tails around these stellar
systems.   Field  and cluster  stars   are  sorted  with  the  help of
color-magnitude diagrams, and star-count  analysis is performed on the
selected cluster stars in order  to increase the signal-to-noise ratio
of their surface density.  We  study the overdensities of these  stars
using the wavelet transform of the star  counts in order to filter the
background  density noise and to  detect the weak structures, at large
scale, formed  by   the  numerous  stars  previously members  of   the
clusters.  We associate these  stellar  overdensities with the   stars
evaporated  from the clusters because   of dynamical relaxation and/or
tidal stripping from the clusters by the galactic gravitational field.
We  take into account the strong   observational biases induced by the
clustering of  galactic field stars and  of background galaxies, along
with the fluctuations of the background due to dust extinction.

Most of the globular clusters in our sample display strong evidence of
tidal interactions with  the galactic plane in  the form of  large and
extended deformations.  These tidal tails exhibit projected directions
preferentially   towards   the galactic   center.    All  the clusters
observed, which   do  not suffer   from  strong observational  biases,
present  such  tidal    tails,   tracing their   dynamical   evolution
(evaporation, tidal  shocking, tidal torquing,  and bulge shocking) in
the Galaxy.  The clusters exhibit different regimes of mass loss rate,
detected  using the  radial density slope  in  the outer parts  of the
clusters.  For NGC~5139  ($\omega$ Centauri), we estimate, taking into
account the possible presence of mass  segregation in its outer parts,
that about 0.6 to  1~\% of its mass  has been lost during the  current
disk shocking  event.  In the  case of NGC~6254,  we tentatively estimate,  in the
cluster reference frame, for  the   radial diffusion velocity  of  the
stars stripped, a value of the order of the velocity dispersion in the
cluster  itself.  The sizes and orientations   of these observed tidal
tails are  perfectly  reproduced by  N-body simulations  of  globular
clusters in the galactic potential well. We present these results in a
companion paper (Combes, Leon \& Meylan 1999).

As a  by-product of this study,  we detect several new galaxy clusters
towards the different fields studied at high galactic latitude. The
estimation of the tidal radius of some of the globular clusters could
have been overestimated because of these galaxy clusters.

               
\end{abstract}

%

\section{Introduction}

Recent works have emphasized that  globular clusters (GCs) are stellar
systems which exhibit strong dynamical evolution in the potential well
of their host galaxy (Gnedin \& Ostriker 1997, hereafter GO97, Murali
\& Weinberg 1997).   Since their birth,  GCs suffer internal evolution
on their own: at the very beginning  GCs evolve rapidly because of the
fast stellar  evolution of their  massive  stars (Vesperini  \& Heggie
1997, Portegies Zwart  \etal\  1997).   The  finite size  of  the  GCs
produces two-body relaxation  between the member  stars; equipartition
of energy heats the lighter stars which diffuse outwards into the halo
while the  heavier  stars  sink slowly  towards the  contracting  core
(Spitzer \& Hart  1971).   Typically,  the relaxation  time   $t_{rh}$
(hereafter we will refer to the relaxation time 
at the half mass radius;  see, e.g., Binney \&  Tremaine 1987) in a GC
is of the order of  $t_{rh} = 10^9$  yr (GO97), which is significantly
less than the age of all galactic globular clusters.

Owing  to  the  negative  specific heat   in  self-gravitating stellar
systems (Antonov  1962,  Lynden-Bell  \& Wood   1968), related  to the
virial   equilibrium, the core is   contracting  monotonically during a
process   called  the  ``gravothermal catastrophe'',   leading to core
collapse,  characterized by very  high stellar densities  (up to $10^6
\mbox{M}_{\odot}\mbox{pc}^{-3}$).  It   has  been eventually   
recognized  that this
process is not so catastrophic after all,  since the cluster core does
not collapse for ever but  bounces back towards lower stellar  density
phases (H\'enon 1975).  The key role of binaries (either primordial or
formed via encounters during the  high stellar density phase) has been
emphasized in the internal dynamics of the GCs:  these binaries act as
a heating source and slow down the core collapse (Goodman \& Hut 1989)
and  even reverse it.  Many  authors have studied numerically the post
core-collapse phase, using conducting-gas-sphere, Fokker-Planck,   and
N-body codes, in simulations   computed  well into core  collapse  and
beyond,   leading  to   the    discovery  of  possible   post-collapse
gravothermal oscillations  (see Meylan \& Heggie  1997  for a review).
Globular clusters   evolve  dynamically, even   when  considering only
relaxation, which causes stars  to escape, consequently cluster  cores
to contract and envelopes to expand.

Any galaxy   through its gravitational   potential well influences the
dynamical  evolution  of its    globular clusters, accelerating  their
destruction.  The stars in the globular cluster halo are stripped by 
the tidal field of
the  galaxy:  the   outward   diffusion of  the    stars   towards the
sub-thermalised halo  is speeded up and the  core contracts  even more
(Spitzer \& Chevalier 1973).  Moreover  the gravitational shocks  heat
up the outer  parts of  the globular  cluster, increasing  its loss of
stars (Aguilar \etal\  1988, Weinberg 1994,  Kundic \& Ostriker  1995,
Leon  \etal\  2000).  Shocks  are  caused by the  tidal  field of the
galaxy: interactions  with the disk, the bulge  and, at a lower level, 
with the
giant molecular clouds  (GMCs,  see Spitzer 1958),  heat  up the outer
regions of  each star clusters.   The disk-shocking occurs when the GC
crosses the    thin disk where   it  is  compressed  by   the  varying
z-component of the  galactic plane potential; this  has been found  to
dominate  the heating of GCs   (Chernoff \etal\ 1986).   A GC globally
gains  energy during the    crossing and exhibits  peculiar  transient
deformation  (Leon \etal\ 2000).  The  shocks with the  bulge and the
GMCs  are   similar in their physical    processes: the GCs  suffer an
elongation aligned parallel to the  density gradient in the bulge
or  the GMC.  These  processes  combined  with the internal  dynamical
evolution  have probably   destroyed  an  important fraction of    the
primordial GCs  and are still currently  at  play.  GO97 estimate that
{\em ``half  of  the present clusters  are to  be destroyed within the
next Hubble time''}.

All  GCs are expected to have   already lost an  important fraction of
their mass, deposited in the  form of individual stars  in the halo of
the Galaxy.  The mass-loss rate is a function of the total mass of the
cluster, its  structural parameters like  the concentration $c$ (where
\conc\  with  \rt\ and \rc\  are  the tidal and  core  radii), and its
orbital  motion around the  galactic center.  Till  recently, the only
way to  investigate the orbital history  of a globular  cluster, apart
from proper  motions, was to derive  its tidal radius \rt, which gives
an indication of its perigalactic  distance.  Unfortunately, there are
difficulties in defining  the   tidal radius, both   theoretically and
observationally; and, as expected, there are discrepancies between the
theoretical and observational values  of the tidal radius (see,  e.g.,
Odenkirchen \etal\ 1997,  Scholz \etal\ 1998).   

N-body  simulations of globular   clusters   embedded in a   realistic
galactic potential (Oh \& Lin 1992; Johnston \etal\ 1997; Combes, Leon
\& Meylan 1999, hereafter CLM99) have been performed in order to study
the amount of  mass loss for different kinds  of orbits  and different
kinds of clusters, along with the dynamics and the mass function in
tidal tails.  The  2-D structure of such tidal  tails appears to be  a
good tracer of  gravitational  shocks and should  be a  tracer  of the
potential well.  Moreover the detection  of unbound stars released  by
the clusters is the only way to measure directly the mass loss rate of
the cluster.

Grillmair \etal\ (1995)  in  an analysis  of star-count in  the  outer
parts of a   few   galactic  globular clusters found     extra-cluster
overdensities that they associated partly with stars stripped into the
Galaxy field.    Similar tidal   interaction remnants around  globular
clusters  have also been found  in external galaxies: Grillmair \etal\
(1996) observed three clusters in M31 which  exhibit departures from a
King profiles.   Leon  \etal\ (1999)  find   tidal extensions  in the
outskirts of interacting binary clusters and  isolated clusters in the
Large Magellanic Cloud (LMC).  Not surprisingly, galactic tidal forces
play also an essential role in the evolution of smaller-scale galactic
clusters,    the  open clusters: they  exhibit   tidal  tails in their
neighborhood (Odenkirchen 1998; Bergond \etal\ 2000).

In this work we study the 2-D structure of  the tidal tails associated
with 20 galactic  globular clusters by using  the wavelet transform to
detect weak structures at large scale and filter the strong background
noise  for the low galactic latitude  clusters.  We analyze with great
care the observational bias which can be very important.  In Section 2
we  present  the  observations, in Section   3  we describe  the  data
reduction, in Section 4 we present the results with, for each globular
cluster, detailed comments about  related observational bias.  Section
5 presents the general discussion of all results.\\

\section{Observations}

\begin{small}
\begin{table}
\begin{tabular}{llcc}
\hline
\hline
Cluster  & Plate/Film   & Filter & Exposure Time                \\
Name	 & ID$^\dagger$ &        & (min)                        \\
\hline
NGC 104  & E12614B  &  R		& 60			\\
	 & E12819B  &  B		& 90			\\
NGC 288  & E12727R  &  R		& 60			\\
	 & E12814B  &  B		& 90			\\
NGC 1261 & ESO155   &  R		& 65			\\
	 & SRC155   &  J		& 65			\\
NGC 1851 & E13033   &  R		& 60			\\
	 & E13030   &  B		& 90			\\
NGC 1904 & E12865   &  R		& 60			\\
	 & E12869   &  B		& 90			\\
NGC 2298 & ESO366   &  R		& 65 			\\
	 & SRC366   &  J 		& 65			\\
NGC 4372 & E13090   &  R		& 60			\\
	 & E13087   &  B 		& 90			\\
NGC 5139 & E12768   &  R	  	& 60			\\
	 & E1228    &  B 		& 90			\\
NGC 5272 & E131	    &  R		&			\\
	 & E131	    &  B 		&			\\
NGC 5694 & ESO512   &  R		& 65			\\
	 & SRC512   &  J		& 65			\\	
NGC 5824 & E12291   &  R		& 60			\\
	 & E12883   &  B 		& 90			\\
NGC 5904 & SRC869   &  R	    	& 65			\\
	 & SRC869   &  J		& 65			\\
NGC 6121 & E12295   &  R		& 60			\\
	 & E12294   &  B		& 90			\\
NGC 6205 & E1069    &  R		&			\\
	 & E1069    &  B		&			\\
NGC 6254 & ESO802   &  R		& 65			\\
	 & SRC802   &  J		& 65			\\
NGC 6397 & E13216   &  R		& 60			\\
	 & E13212   &  B		& 90			\\
NGC 6809 & E12447   &  R		& 60			\\
	 & E13195   &  B		& 90			\\	   
NGC 7492 & SRC676   &  R		& 65			\\
	 & SRC676   &  J		& 65			\\
Pal 5	 & SRC869   &  R		& 65			\\
	 & SRC869   &  J		& 65			\\
Pal 12	 & ESO600   &  R		& 65			\\
	 & SRC600   &  J		& 65			\\
\hline
\end{tabular}

$^\dagger$ Northern GCs are taken on POSS~I plates.
\caption{List of  observed globular clusters.  About plates/films, ESO
and  SRC stand for  the plates of  the corresponding two surveys.  All
the other  plates/films are   from  our observations, except  the  two
plates E131 and E1069 which are from the POSS~I.}
\label{table_observation}
\end{table}
\end{small}

We chose  to observe, for  the present  study,  a  number of  clusters
sharing different properties or  locations in the Galaxy, with various
masses  and structural parameters.   In  order  to  detect the   tidal
extensions around globular clusters, it is necessary to have very wide
field  observations.    During the   years    1996 and  1997,  optical
observations    were performed at   the ESO    Schmidt telescope  with
photographic films.  The   field of  view   is of  $5.5\degree  \times
5.5\degree$  with a scale of  67.5\arcsec/mm.  In combination with the
UG1, BG12 and RG630 filters, the Kodak Tech Pan emulsion 4415 provides
passbands  close to $U$, $B$,  and $R$,  respectively.  Eventually, we
use only the $R$ and $B$ filters for this analysis, since the $U$ band
images are  significantly shallower than  the images in  the other two
filters.  The sample  of clusters is enlarged by   using some ESO  and
SERC survey plates available at the Centre  d'Analyse des Images (CAI)
of the  Paris  Observatory, along  with  a  few POSS~I  plates  kindly
provided by  Leiden  Observatory.  All the  observed globular clusters
studied in this paper are listed in Table \ref{table_observation} with
corresponding  plates/films  identification   numbers,  filters,   and
exposure  times.  All     these photographic  plates   and films   are
digitized using  the  MAMA  (Machine Automatique  \`a   Mesurer pour
l'Astronomie) scanning machine at the CAI, which provides a pixel size
of  10  \micron.  The  astrometric  performances   of the machine  are
described in Berger \etal\ (1991).

Table   \ref{table_sample}   displays  for   each  cluster, its  J2000
equatorial  coordinates ($\alpha$,$\delta$),  its galactic coordinates
(l,b),   its  equatorial rectangular coordinates    (X,Y,Z), and, when
available,   its velocity   components    (U,V,W) in   the  equatorial
rectangular system.

\begin{table*}
\begin{tabular}{lccrrrrrrrr}
\hline
\hline
Cluster  &	 $\alpha$ & $\delta$ &  l~~	& b~~ & X~~	&	Y~~	& 	Z~~	& U~~	& V~~	& W~~  \\
Name & (J2000)	& (J2000) & (\degree)~~ & (\degree)~~ & (kpc) & 	(kpc)   & 	(kpc)	& (km/s) & 
(km/s) & (km/s) \\
\hline
NGC 104 & 00 24 05.2 & --72 04 51 &  305.9 & --44.9 &1.7	&	--2.4	&	
--2.9	&--91	& 169 	& 33\\
NGC 288$^\dagger$ & 00 52 47.5 & --26 35 24  & 152.3 & --89.4& --0.1	&	
0.0	&	--8.1	& --10.	& --39	& 53\\
NGC 1261 &  03 12 15.3 &  --55 13 01 & 270.5 & --52.1 & 0.1	&	--9.3	&	
--12.0	&	&	& \\
NGC 1851$^\dagger$& 05 14 06.3 & --40 02 50 & 244.5 & --35.0  &  --4.1	& 	--8.6	&
	--6.7	&--234	& --37   & --107\\	
NGC 1904&05 24 10.6 & --24 31 27 & 227.2 & --29.4 & --7.2	&	--7.8	&
	--6.0	&	&	&\\
NGC 2298& 06 48 59.2 & --36 00 19 & 245.6  &--16.0& --4.1	&	--9.1	&	--2.9
	&	&	&\\
NGC 4372& 12 25 45.4 & --72 39 33 & 301.0  & --9.9 &2.3	&	--3.9	&	--0.8
	&	&	&\\
NGC 5139& 13 26 45.9&  --47 28 37&  309.1 &  15.0 &3.0	& 	--3.7	&	1.3
	& 120	& --13	& --87\\
NGC 5272&13 42 11.2&  +28 22 32 &  42.2  & 78.7 & 1.4	&	1.3	&	9.5
	&--55	& 44	& --106\\
NGC 5694&14 39 36.5 & --26 32 18 & 331.1  & 30.4&  24.9	&	--13.8	&	16.7
	&	&	&\\
NGC 5824 &15 03 58.5 & --33 04 04 & 332.6 &  22.1& 26.4	&	--13.7	&	12.1	&	& & \\
NGC 5904&15 18 33.8 & +02 04 58  &  3.9 &  46.8& 4.8	&	0.3	&	5.1
	&317	& 195	& --202\\
NGC 6205&16 41 41.5 & +36 27 37 &  59.0 &  40.9 & 2.6	&	4.4	&	4.4	
&--260	& 145	& --114\\
NGC 6254& 16 57 08.9 & --04 05 58 &  15.1 &  23.1 &3.7	&	1.0	&	1.6
	& 80	& --11	& 197 \\
NGC 6397& 17 40 41.3&  --53 40 25 & 338.2&  --12.0 &2.0	&	--0.8	&
	--0.5	&--38	& 131 	& --103\\
NGC 6535&18 03 50.7 & --00 17 49 &  27.2 &  10.4 & 5.8	&	3.0	&	1.2
	&	&	&\\
NGC 6809& 19 39 59.4 & --30 57 44  &  8.8 & --23.3 &4.6	&	0.7	&	--2.0	&
	&	&\\
NGC 7492& 23 08 26.7 & --15 36 41  & 53.4 & --63.5 &6.5	&	8.7	&	--21.8	&
	&	&\\
Pal 5$^\ddagger$  & 15 16 05.3 & --00 06 41  &   0.9 &  45.9	& 15.2	&	0.2	&
	15.7	& 50	& --47 	& --108\\
Pal 12	& 21 46 38.8 & --21 15 03 &  30.5 &  --47.7 &10.3	&	6.1	&	--13.2	&\\
\hline
\end{tabular}
\caption{The  X,Y,Z positions  are from  Harris  (1996) and the  U,V,W
velocities from Dauphole  \etal\  (1996).   ~~$^\dagger$  These cluster
velocities   are  from Dinescu   \etal\  (1997).   ~~$^\ddagger$ These
cluster velocities are from Scholz \etal\ (1998).}
\label{table_sample}
\end{table*}

\section{Data reduction}

\subsection{Source extraction}
Once  the plates/films are digitized,   the  next step consists   of
identifying all point sources in  these frames.  The source extraction
is performed on each frame  using SExtractor (Bertin \& Arnouts 1996),
a software dedicated to  the automatic analysis of astronomical images
using a   multi-threshold algorithm  allowing  good object deblending.
The detection of the stars  is done at   a 3-$\sigma$ level above  the
background. This software, which can deal with huge amount of data (up
to 60,000 $\times$ 60,000 pixels) is not suited for very crowded field
like  the centers of the  globular clusters.  Since the radial surface
density is so much unreliable towards the very  crowded parts of these
globular clusters, we just ignore it in all crowded inner areas.  From
the catalogues in $B$ and $R$ filters, produced by SExtractor for each
cluster, we construct a color $B$ and color index $B-V$ catalogue with
the instrumental magnitudes.  We do not  calibrate our data, except in
the case of   NGC~5139, since we  need  only relative magnitudes   and
colors for the  purpose   of  establishing cluster membership.     The
magnitude error from  the photographic plate is  found to be up to 0.2
mag for  the faintest  stars.   Typically, we get,  for each  field, a
total number of stars from $7 \times 10^4$ up  to $ 2 \times 10^6$ for
the richest fields.  We do not  apply  any crowding  correction to our
stellar counts, first, because crowding is nearly constant and weak in
the   outer areas  (the  only ones  we  consider)   surrounding of the
clusters (see also   Grillmair   \etal\ 1995), and,  second,   because
crowding is completely dominated  by the observational biases for  the
overdensities.
 
\subsection{Star/Galaxy separation}

\begin{figure}[hbpt]
\caption{Star/galaxy magnitude  vs.  log(star/galaxy area) diagram for
the field around  NGC~288. At large  areas and bright magnitudes,  the
upper sequence is made  of stars, while  the lower sequence is made of
galaxies. The lowest line represents  the  3-$\sigma$ limit below  the
continuous line, fitting the star sequence.}
\label{fig_separation_gal_star}
\end{figure}

In  case of low fore-   and background densities towards a  considered
globular cluster, i.e., for GCs located at high Galactic latitudes, we
perform a star/galaxy separation  by  using the method of  star/galaxy
magnitude vs. log(star/galaxy area) - which is shown to work well down
to     the     18  instrumental      magnitude,  as     displayed   in
\fig{fig_separation_gal_star}.    The  galaxies have  a  lower surface
brightness than  the stars and  in the magnitude vs.  log(area) plane,
the  two classes of objects follow  different loci as clearly shown on
\fig{fig_separation_gal_star}.  The star sequence is fitted by a fifth
or sixth
order polynomial and  the objects more than 3~$\sigma$ fainter  than 
the fit are
considered as  galaxies, through an iterative  process which  stops as
soon as no   new  galaxies are  detected  (typically  after  about  10
iterations).

\begin{figure}[hbpt]
\caption{Overdensities  of   the  galaxies (plate   SRC678)  using the
spatial details down to the plane 2  of the WT.  North  is up, East to
the left.  The Abell clusters in  this field are labeled. The contours
levels represent 3-$\sigma$ to 8-$\sigma$  detections.  The voids near
the Eastern corners are due to artifacts from this plate.}
\label{fig_cluster_galaxie}
\end{figure}

We detect some background clusters  of galaxies from the overdensities
at high   spatial frequencies, using the   Wavelet Transform  (WT, cf.
Slezak \etal\  1994 and hereafter).  As a check, the  galaxy catalogue obtained with
this method  is  shown  to correlate   very well,  with  the
 Abell cluster catalog (Abell \etal\  1989).  We present in
\fig{fig_cluster_galaxie} the detection of  clusters of galaxies  in a
wide field observed with the plate SRC678: the use  of the WT allows a
clear detection of about 50 clusters and  substructures of clusters of
galaxies at the typical scale of these structures.  We point  out that 
we refer   to Abell cluster  without any
distinction between north and south Abell  clusters. 
We emphasize that
the star catalogues produced from such  galaxy/star separation
is    polluted by galaxies    at   faint  magnitudes,  as visible   in
\fig{fig_cluster_galaxie} because of confusion at faint magnitude 
for the detection.  Our problem being the genuine detection of
star  overdensities, we look   at  the correlation between  the galaxy
clusters detected and the star overdensities,  in order to disentangle
the  confusion, assuming that the  faint magnitude galaxies follow the
distribution of   the bright galaxies  detected.    This assumption is
justified by comparing  some fields heavily  polluted by galaxies with
the so-called {\em  stellar} overdensities  (see, e.g., \fig{fig_pal5}
hereafter). 
In order to remove the bad classification on the globular cluster region
because of the crowding, we set the surface density of this area ($r < r_t$)
to zero. 

\subsection{Star selection}

Following the method of  Grillmair \etal\ (1995)  we perform, for each
considered globular cluster, a star selection from the color-magnitude
diagram (CMD) in which cluster stars and field stars exhibit different
colors.  In this way we   can differentiate present and past   cluster
members  from the fore- and  background  field stars by identifying in
the CMD the area occupied by cluster stars.  The envelope of this area
is empirically chosen so as to optimize the  ratio of cluster stars to
field stars in the relatively sparsely populated outer regions of each
cluster.

\begin{figure}[hbpt]
\caption{Upper right  panel :  color-magnitude
diagram of stars in the cluster NGC~2298 (r $< 0.5r_t$) using instrumental
magnitude. Upper left panel :  color-magnitude
diagram of stars in the cluster field (for clarity only 10 \% of the total stars 
is shown).
Lower panel : Signal/Noise $\tilde{s}$   distribution  (see text)   in  the
color-magnitude diagram (CMD)  built  from the  NGC~2298 plates.   The
resolution  used with  the WT  takes  into account  the error  on  the
magnitude  ($\sim$  0.2   mag).   Values greater    than  zero  (dark
isopleths) indicate greater contributions  from the cluster stars than
from the background stars (see Equ.\ref{equa_sn}).}
\label{fig_cmd_sn}
\end{figure}

This  search  for a mask  is  done  by subdividing the color-magnitude
plane into a  50 $\times$ 50 array in  which individual  sub-areas are
about 0.1 mag wide in color and 0.15 mag high in $B$ mag  depending slightly
on the cluster.  Assuming that the
color-magnitude distribution of the field stars is constant across the
plate,  a color-magnitude sequence  for each  cluster can be estimated
from:
\begin{equation}
f_{cl}(i,j)=n_{cl}(i,j)-gn_{f}(i,j)
\end{equation}
in the notation of Grillmair \etal\ (1995) where $n_{cl}$ and $n_f$ 
refer to the number of stars with color index
i and the instrumental magnitude  index j  counted within the  central
region  of the   cluster  and  in   an annulus outside   the  cluster,
respectively.  The factor  g is the ratio of  the area of  the cluster
annulus to  the  field annulus.   We compute  the  ``signal-to-noise''
ratio for each color-magnitude sub-area:
\begin{equation}
s(i,j)\equiv \frac{f_{cl}(i,j)}{\sqrt{n_{cl}(i,j)+g^2n_f(i,j).}}
\label{equa_sn}
\end{equation}
Given the magnitude resolution  of about 0.2 mag,  for both plates and
films, we perform a wavelet transform  (WT, cf. Slezak \etal\ 1994)
of  the    s(i,j)     function     on   5 planes,     with
$W(l,i,j)=WT(s(i,j))$. We remove the planes 0 and 1  to obtain the S/N
function 
\begin{equation}
 \tilde{s}(i,j)=\sum_{l=2}^{5}W(l,i,j)
\label{equa_sntilde}
\end{equation}
with
a magnitude resolution of at most 0.2 mag, see below for more details
on the WT.

Grillmair  \etal\ (1995) estimated a  CMD  mask  for the selection of
stars by computing a cumulative function from the  S/N function for each cluster.  
This cumulative function  reaches a     maximum  for a  sub-area   of   the
color-magnitude   plane.   Then by  selecting   all sub-areas with S/N
values higher than this maximum, it is possible to construct the mask.
In the  present case, we depart slightly  from Grillmair \etal\ (1995)
procedure by selecting, for some fields, a subset  of the mask, with a
higher  S/N   value     relative   to  the background      stars  (see
\fig{fig_cmd_sn} and \fig{fig_cmd_tot}).  It must be a  compromise 
between the S/N ratio and
the number of stars selected,  in order  to  get a sufficient  spatial
resolution which is lowered by Poissonian noise for small star counts.
Given the S/N chosen, we have been able to eliminate from 50~\% up 
to 99~\% of the field stars. We show in \fig{fig_cmd_tot} the CMD selected
for the less contaminated fields.

\begin{figure*}[hbpt]
\caption{Color magnitude diagrams (left panel), using instrumental magnitude, of
stars in the fields of 8  clusters with
the area selected from the highest contrast between the cluster and
the field. Note that the range can be different for each cluster.  The right
panel shows the S/N distribution $\tilde{s}$ which gives the best contrast for the selection
in the CMD space (see Equ.\ref{equa_sntilde}).
For the constrast a darker color means a greater contribution from the cluster stars.
In the case of NGC~288, there is no smoothing. Note that each CMD has been scaled for
matching the $\tilde{s}$ map. }
\label{fig_cmd_tot}
\end{figure*}

On the CMD-selected  star-count map $M(x,y)$, we  fit a background map
$Z(x,y)$, following Grillmair \etal\ (1995), by masking the GC (1 to 2
$r_t$) and using  a blanking value inside,  equal to the  mean between
1.5  and 2.5  $r_t$ to get  a smooth  background.  We  fit, on a 
$128\times 128$ binned grid, a low-order
bivariate  polynomial surface $Z(x,y)$,  mainly first- or second-order
surface, to avoid   to  erase some  local  variation:
\begin{eqnarray}
Z(x,y) & = & \sum_{i,j}a_{ij}x^{i}y^{j} 
\quad \mbox{with} \quad  0 \le i,j\le 2.
\end{eqnarray}
We  subtract this background from the   CMD-selected map to  get a
surface-density map $T_r(x,y)$ of the overdensities that we can attribute to the
tidal extension of the GC:
\begin{eqnarray}
T_r(x,y) &  = &  M(x,y)-Z(x,y) 
\end{eqnarray}
after having  analyzed the potential  observational biases  that could
create the fluctuations in the star-counting analysis.

\subsection{Wavelet analysis}
The wavelet transform is a powerful signal processing technique which provides
a decomposition of the signal into elementary local contribution labeled 
by a scale parameter (Grossman \& Morlet 1985). They are the scalar products
with a family of shifted and dilated functions of constant shape called
wavelets. The data are unfolded in a space-scale representation which is
invariant with respect to dilation  of the signal. Such an analysis is 
particularly suited to study signals which exhibit space-scale discontinuities 
and/or hierarchical features. Its ability to detect structures at particular 
scales has already been used in  several astrophysical problems 
(Gill \& Henriksen 1990, Slezak \etal\ 1994, Cambr\'ezy, L. 1999, Chereul et al. 1999)

\subsubsection{``A trous'' algorithm}
We perform  on the raw tidal  map $T_r(x,y)$ a wavelet analysis using
the ``\`a trous'' algorithm (see  Bijaoui 1991).   It  allows to get  a
discrete wavelet decomposition   within  a reasonable CPU time.    The
kernel function  $B_s(x,y)$  for the  convolution   is a $B_3$  spline
function.  The  wavelet decomposition $W(i,x,y)$  is obtained from the
following steps:
\begin{eqnarray}
c_o(x,y) & = & image(x,y), \\
c_i(x,y)& = & c_{i-1} \convolution B_s(\frac{x}{2^i},\frac{y}{2^i}), \\
W(i,x,y)& = & c_i(x,y)-c_{i-1}(x,y).  
\end{eqnarray}
The last plane, called Last Smoothed Plane  (LSP), is the residuals of
the last convolution  and not a wavelet  plane, but afterwards we will
abusively  speak of wavelet  plane for  all  these planes. Each  plane
$W(i,x,y)$ represents  the details of the  image at the   scale i.  We
divide each image in $128\times 128$ bins, a process which changes the
spatial resolution of each cluster according to the different sizes of
the   fields:  typically  we  get  star-count    maps of  3\arcmin\ to
16\arcmin\ resolution. The spatial resolution $\sigma_\theta$ for each
wavelet-rebuilt cluster  field can  be  computed, in  arcmin, from the
following   relation:   $\sigma_\theta =  0.0538\theta_{field}$,  with
$\theta_{field}$ being  the total field size  in arcmin.  It is so far
possible to do a  filtering  of each plane  to  get only the  relevant
wavelet  component.  One problem  is to find the  noise level for each
plane.  We know that  the raw tidal map  is blurred by the  Poissonian
noise of  the   background objects; this   is especially  true at  low
galactic latitudes.    We  could  perform an  Anscombe  transformation
(Murtagh  \etal\ 1995) to   transform   the Poissonian  noise into   a
Gaussian  noise on   each  scale.   Actually  we  choose   to  perform
Monte-Carlo simulations,   because  of the varying    Poissonian noise
through the field, in order to follow  easily the spatial variation of
the rms noise at each scale. The contours for the surface density are
computed to be above 3 $\sigma$ level, with $\sigma$ being the rms 
fluctuation of the {\it selected} wavelet coefficients computed in an
area avoiding the central cluster.

\subsubsection{Filtering of high varying density background noise}
\begin{figure}[hbpt]
\caption{Filtered   image of  color-selected star-count  overdensities
(Log) in NGC 5139 using the Wavelet Transform (WT) (upper panel) to be
compared  with the  raw  star-count (lower  panel).   The upper  panel
displays the full resolution of $3.2^\prime$ using the whole set of wavelet planes.}
\label{fig_filtering}
\end{figure}

In  case of strong  gradient density of  the  galactic background, the
noise is varying with the  location in the  field.  To filter properly
this    noise, we perform N   Poissonian  simulations  from the fitted
background  star  counts   $Z(x,y)$ and   we take   the WT  of   the N
realizations and perform statistics on each pixel for the whole set of
wavelet planes, 
\begin{eqnarray*}
W_{n}(i,x,y) & = & WT(Poisson(Z(x,y))) \quad \mbox{with} \quad  n=1,N. \\
\end{eqnarray*}

Practically we have taken N=100. Then we  fit by a low-order  
bivariate polynomial surface  a rms noise
map  ($\sigma_{bck}(i,x,y)$), for  each wavelet  plane,  to obtain  an
estimate of the rms fluctuation on the N  realizations.  This allows a
good estimate of the rms noise without  the need of performing a great
number of simulations, which are CPU time-consuming because of the WT:
\begin{displaymath}
\sigma_{bck}(i,x,y)=\sum_{k,l}a_{k,l}^{(i)}x^{k}y^{l} \quad \mbox{with} \quad 0 \le k,l \le 2 \\
\end{displaymath}

We filter each   component of the  raw  map $W(i,x,y)$, above a  given
threshold   $\alpha$, using the rms   noise  $\sigma_{bck}(i,x,y)$ map on  each
wavelet scale  $i$ to get the filtered  wavelet planes $W_f(i,x,y)$ of
our image:
\begin{eqnarray*}
W_f(i,x,y) & = & W(i,x,y)  \mbox{\small \hspace{0.1cm}  
                 If $|W(i,x,y)| > \alpha  \sigma_{bck}(i,x,y)$} \\
           & = & 0         \mbox{\small  \hspace{1.4cm} Otherwise}       
\end{eqnarray*}

In this study the coefficients are filtered  at the 3-sigma level.  We
show the  case  of the   globular  cluster  NGC~5139,  located at  low
galactic  latitude, for which we  present the  raw surface density map
and the filtered map  at different resolution (see \fig{fig_filtering}
and  \fig{fig_resolution}).  We remind that  a wavelet plane $i$ has a
typical    resolution  of  $0.86\times2^i$     pixels,  which  is  the
Gaussian-equivalent resolution  of  the wavelet function at  the scale
$i$.  We have to point out that our filtered solution is {\em not} the
optimum  solution  (cf.  Starck \etal\   1997,  e.g.  for 1-D  optimum
solution) since  it is  only a  selection of  significant coefficients
from the raw map.  
The ``\`a trous'' algorithm implemented  permits a WT transform with a
rate of about 2 kPixel/s/plane on a DecAlpha500 workstation 
\footnote{The~IDL~and~C~procedures~for~the~``A trous''~algorithm   are
available at {\tt
ftp://smart.asiaa.sinica.edu.tw/pub/sleon/wavelet.tar.gz} or under
request to sleon@asiaa.sinica.edu.tw}
 
\begin{figure}[hbpt]
\caption{Different resolutions  of the  tidal tail extensions  towards
NGC~5139 using the WT   filtered planes. The spatial   resolutions are
3.2\arcmin,     6.4\arcmin, 12.9\arcmin, 25.8\arcmin, 51.6\arcmin, and
103.2\arcmin\ from the upper-left panel to the lower-right panel.}
\label{fig_resolution}
\end{figure}

The final tidal map $T_f(x,y)$ is built as follows:
\begin{equation}
T_f(x,y)=\sum_i W_f(i,x,y) \quad \mbox{with}\quad  i_1\le i \le i_2 
\end{equation}
where $W_f$  is the filtered  WT in case   of strong background noise.
The lower and upper  indexes $i_1$ and  $i_2$ constrain the resolution
of  the  rebuilt map  by filtering  the higher and/or  the lower space
scale wavelet planes.   For the adopted binning, we  find that the map
with the planes 3  to 7 (LSP) gives  the  best compromise between  the
spatial resolution and the Poissonian noise of the star-counting after
the filtering of the background noise.  It provides,  in most cases, a
higher spatial resolution than in Grillmair \etal\ (1995), because the
wavelet decomposition extract the energy only at the useful scales. We
have  to  point  out  nevertheless  that  the  wavelet   basis used
here  is not orthogonal, mixing up slightly the scale energy on different
planes, but this does not affect our rebuilt map.

\section{Results}
In this section we present all results  related to the observations of
stars surroundings  of each GC.   We discuss individually each cluster
for    the particular observational  biases   which  could  affect its
results.   Grillmair \etal\ (1995) found that the clusters in their sample with 
obvious tidal extensions showed a break in their surface density profiles, 
becoming pure power law at large radii. We   try to link  in  a  systematic 
way the   shape of each
observed tidal tail to the orbital phase of the corresponding cluster.
For this we define \q{a}{b} as the slope of the radial surface density
between   $a\times\rt$ and $b\times\rt$.    The  slope  of the  radial
surface density is computed when the  tidal tails are not dominated by
the noise which would lead to a flat slope.  In practice, we choose to
compute the three slopes \q{1}{3}, \q{3}{6}, and \q{1}{6}, and give in
\tab{slope_profile} their   values   only   for clusters    where  the
signal/noise  ratio   for   these  azimuth   averaged  parameters   is
sufficiently high.     Practically,  this means  that   we  remove all
clusters with \q{1}{3} shallower than -0.5.
Whenever possible,  our surface density  profiles \footnote{The 1-d and 
2-d star counts are available by request to the authors.}
are extended inwards
with  the  surface-brightness profiles   from  Trager   \etal\ (1995),
assuming a linear  relation between  light emission  and stellar surface 
density through the globular cluster.  Crowding and saturation problems
in  our plates/films make  the inner  parts  of  our  density profiles
highly  unreliable.   Consequently, the   adjustment  between Trager
\etal 's profiles  and ours is  done, in the  short radius range where
both profiles are reliable, by adjusting a constant
$K$ in the following way:
\begin{equation}
\log(\mbox{surface density})=-\mu/2.5 + K
\end{equation}
where $\mu$  is the fitted surface brightness  at r (see Trager \etal,
1995). For Trager \etal\ (1995) profiles, only data outside the radius
r=1\arcmin\ are shown.  We point out that differences between the two
profiles  in the  very outer parts   can partly  be explained by  mass
segregation in the cluster, unveiled by different limiting magnitudes.

\begin{table}
\begin{tabular}{lccc}
\hline
\hline
Cluster Name &  \q{1}{3} & \q{3}{6} & \q{1}{6} \\
\hline

NGC 288 & -0.90 (0.30) & -1.56 (0.90) & -1.18 (0.18) \\
NGC 1261& -0.96 (0.31) & 0.08  (0.20) & -0.64 (0.12) \\
NGC 1851& -1.84 (0.20) & -0.47 (0.21) & -0.98 (0.14) \\
NGC 1904& -1.01 (0.17) & -0.25 (0.25) & -0.65 (0.09) \\
NGC 2298& -1.92 (0.77) & -3.41 (0.96) & -1.43 (0.39) \\
NGC 5139$^\dagger$& -5.03 (0.32) &	      &		     \\
NGC 5272& -0.62 (0.28) & -0.40 (0.55) & -0.35 (0.24) \\
NGC 5824$^\ddagger$& -3.16 (0.64) & 	      &		     \\
NGC 6254& -1.91 (0.22) & 	      & 	     \\
NGC 6535& -1.02 (0.35) & 	      &	-1.69 (0.56)	     \\
NGC 6809& -0.96 (0.25) &	      &		     \\
Average$^\S$ & -1.24 (0.5) &  -1.00 (1.3) & -0.91 (0.37) \\
\hline
\end{tabular}

$^\dagger$  \q{1}{2}\\
$^\ddagger$ Fit between 10 et 20\arcmin \\
$^\S$ Statistical dispersion (NGC~5139 not included).
\caption{Slopes \q{a}{b} of   the radial surface density  profiles for
different   ranges  in radius for  some   the globular clusters in our
sample (see text for explanations).   We  consider here only  clusters
with \q{1}{3} smaller than -0.5. Note that the error bars do not include the 
uncertainties in the background determination.}
\label{slope_profile}
\end{table}

It is worth mentioning that the measured slope will be flattened at 
small radii since
the closer to   the  cluster  the larger  the  crowding   and  since a
azimuthal averaged  value is more  sensitive to noise  at large radii.
For a power  law dependence, with a slope  $\alpha$, of the tidal tail
surface   density,  the tail/noise  surface   density ratio  scales as
${r^{-\alpha}}/{\sigma_{bck}}$, where $\sigma_{bck}$ is the background
surface density.  We choose the quantity \q{1}{3} ---  the only one we
are able to   determine  in a large enough   number  of GCs ---  as  a
quantitative estimator for  comparing  the outer structures  of  these
clusters.  Table \ref{table_dynamic} gives, for all GCs in our sample,
the dynamical and  structural   parameters (from GO97 and   references
therein) which  are  representative  of  the internal   and   external
dynamical evolution of   these  globular  clusters; $t_{rh}$ is    the
relaxation   time  at  the  half-mass   radius;   $\nu_{evap}$ is  the
destruction rate  due to evaporation; the ratio $\nu_{tot}/\nu_{evap}$
of the   total destruction  rate   to  the   destruction rate  due  to
evaporation illustrates the  importance of the galaxy-driven evolution
suffered  by the clusters;  $c=\log(\rt/\rc)$ is  the concentration of
the cluster, where   \rc\  and \rt\  are   the core and  tidal  radii,
respectively; M is the cluster mass and  $V_{HB}$ is the $V$ magnitude
of the horizontal branch stars. Given the low surface density and low
S/N of some  cluster tidal tails, the radial surface density profile 
is not shown for all the clusters.
 
\begin{table*}
\begin{tabular}{lccccccccc}
\hline
\hline
Cluster	& $t_{rh}$ & $\nu_{evap}$ & $\nu_{tot}/\nu_{evap}$ & c	& $r_c$	&  $r_t$ & M	& $V_{HB}$ &Ref.\\
Name	& ($10^9$yrs)	&	&	&	& (pc)	&(pc)	& ($10^5\msun$)	&  & \\
\hline
NGC 104	& 3.94	& 0.07	& 2.1	&	2.04	&	0.50	&	54.8	& 14.5	& 14.06	& 1\\
NGC 288	& 1.42	& 0.37	& 3.0	&	0.96	&	3.47	&	31.6	& 1.1	& 15.38	& 1  \\	
NGC 1261& 1.51	& 0.23 	& 1.2	&	1.27	&	1.82	&	33.9	& 3.3	& 16.70	& 2\\
NGC 1851& 2.72	& 0.12	& 1.0	&	2.24	&	0.28	&	48.7	& 5.6	& 16.15	& 1\\
NGC 1904& 0.88	& 0.28	& 1.3	&	1.72	&	0.60	&	31.5	& 3.6	& 16.15	&1\\
NGC 2298& 0.34	& 1.02	& 1.0	&	1.40	&	1.00	&	19.1	& 0.7	& 16.11	&3\\
NGC 4372& 2.76	& 0.12	& 3.8	&	1.30	&	2.63	& 	52.5	& 3.2	& 15.30	&2\\
NGC 5139& 10.1	& 0.04	& 9.4	&	1.24	&	3.72	&	64.6	& 51.0	& 14.53	&1,8\\
NGC 5272& 7.28	& 0.03	& 1.6	&	1.85	&	1.48	& 	104.8	& 7.8	& 15.65	& 1\\
NGC 5694& 1.17	& 0.21	& 1.0	&	1.84	&	0.59	& 	40.8	& 2.9	& 18.50	& 1\\
NGC 5824& 21.3	& 0.02	& 1.6	&	2.45	&	0.52	&	146.6	& 9.0 	& 18.60	& 1\\
NGC 5904& 3.75	& 0.07	& 26.0$^\dagger$&	1.87	&	0.89	&	66.0	& 8.3 & 15.06& 1\\
NGC 6205& 3.17	& 0.09	& 1.1	&	1.49	&	1.82	&	56.2	& 6.3	& 14.90	& 1\\
NGC 6254& 0.76	& 0.40	& 1.1	&	1.40	&	1.07	&	25.7	& 4.9	& 14.65	 & 1\\
NGC 6397& 0.18	& 2.35	& 1.1	&	2.50	&	0.03	&	66.0	& 1.6	& 12.87	& 4\\
NGC 6535& 0.25	& 1.37	& 1.4	&	1.30	&	0.83	&	16.6	& 0.6	& 15.73	& 1\\
NGC 6809& 1.64	& 0.58	& 1.2	&	0.76	&	3.98	&	22.9	& 2.4	& 14.40	& 1\\
NGC 7492& 2.75	& 0.19	& 77.8$^\dagger$&	1.00	&	6.17	& 	61.7	& 0.6	& 17.63	& 5\\
Pal 5	& 7.17	& 0.16	& 1.3	&	0.74	&	19.5	&	107.2	& 0.3	& 17.51	& 6\\
Pal 12	& 1.52	& 0.40	& 17.9$^\dagger$&	0.90	&	6.17	&	49.0	& 0.2	& 17.13	& 7\\
\hline
\end{tabular}
\caption{Dynamical and structural parameters  linked to  the dynamical
evolution of  the globular  clusters in our   sample  (from Gnedin  \&
Ostriker 1997, GO97).  The  destruction rate $\nu_{tot}$ includes  the
total destruction rate due to disk and bulge shocks  from the model by
Bahcall \etal\ (1983).  The structural properties of the GCs come from
the  following   references: (1) Pryor \&   Meylan  (1993), (2) Hesser
\etal\ (1986),  (3)  Geisler \etal\  (1995), (4) Meylan \& Mayor (1991),
(5)   Webbink (1981), (6)
Schweitzer \etal\  (1993), (7) Armandroff \& Da  Costa (1991), and (8)
Meylan   \etal\ (1995).  All $V_{HB}$   values are from Harris (1996).
~$^\dagger$  These high ratio values    are due to the  gravitational
shocks, stronger  in the Bahcall    \etal\ (1983) model than  in   the
Caldwell \& Ostriker (1983) model (see GO97).}
\label{table_dynamic}  
\end{table*}

\subsection{NGC~104 $\equiv$ 47~Tucanae}

\begin{figure}[hbpt]
\caption{NGC~104 $\equiv$  47~Tuc.  ~~(a): Surface density
plot  displaying  tidal tails (in   Log) around NGC~104 (47~Tuc).  The
different arrows indicate the directions of  the cluster proper motion
(dotted arrow), of the galactic   center (dashed  arrow), and of   the
direction  perpendicular  to the  galactic   plane (solid arrow).  The
dashed circle centered on the cluster indicates its tidal radius.  The
horizontal  double  arrow stands for  150~pc.   (b): IRAS
100-\micron\ chart overlaid  with the above tidal-tail surface density
contours.  The strong emission  in the S-E  corner is coming  from the
Small Magellanic Cloud (SMC).}
\label{fig_n104}
\end{figure}

NGC~104 is at a distance of 4.1~kpc from the  sun, with its horizontal
branch (HB) at $V$ = 14.06~mag.  It has a  tidal radius of about 55~pc
with a   rather      high   concentration  \conc\    =    2.04    (see
\tab{table_dynamic}).  It is one of  the most  massive and nearby  GCs
(Meylan \& Mayor 1986,   Meylan  1989).  In their   study, Odenkirchen
\etal\ (1997) estimate that   NGC~104 has experienced a  disk crossing
about 60~Myr ago and  suffers   frequent disk-crossing events with   a
period of  about 160 Myr between  each passages.  The rotation of this
cluster (Meylan \& Mayor 1986) should enhance its  mass-loss rate by a
factor of about 1.5  (Longaretti \& Lagoute 1996).   Unfortunately the
detection  of extra-tidal material   is made  difficult by the  strong
pollution along  the line of sight  due to the Small  Magellanic Cloud
(SMC) background stars:  in the CMD  of NGC~104, the cluster sequences
are superposed, with a  vertical translation, with the sequences drawn
by the   stellar populations in  the  SMC.  In  order  to discriminate
efficiently  between   cluster   and   SMC stars,    we   compute  the
tail/background~ S/N  function relative   to the  SMC  by  taking  the
background  field  for  the  s(i,j) function  on  the  SMC (see
\fig{fig_cmd_tot}.  
We fit  a 3$\times$3     bivariate   surface in    order    to reproduce the SMC
overdensity. A strong  residual of the  SMC is still present (see Fig.
\ref{fig_n104}), fortunately located in the S-E corner.  It represents
a high surface density  of stars of the  SMC, well correlated with the
dust emission seen  in the IRAS  100-\micron map.  About the  globular
cluster itself,  tidal extensions towards the  N-W and  S-W directions
are marginally present around  NGC~104.  The IRAS 100-\micron map does
not  exhibit any anticorrelation  with these tidal extensions.  No fit
of the surface density profile of these tidal tails has been performed
because of their poor statistical significance. A steep mass function
must be present in the tidal tails because of the mass segregation
observed in this cluster (Anderson \& King, 1996)

\subsection{NGC~288}

NGC~288 is at a distance of 8.1~kpc  from the sun, with its horizontal
branch (HB) at $V$ = 15.38~mag.  It has a tidal  radius of about 32~pc
(see \tab{table_dynamic}).  Its  concentration is  low, with \conc\  =
0.96.  It is  located close to the South  Galactic Pole, at 8 kpc from
the Sun (Harris 1996), with  a retrograde 
orbit (Dinescu \etal\ 1997). From GO97, NGC~288  is a cluster  
with a dynamical evolution  strongly
driven by  the galactic tidal   field (see Table \ref{table_dynamic}).
NGC~288 was already observed  by  Grillmair \etal\ (1995),  who  found
tidal extensions  on a field $200\arcmin\times200\arcmin$ smaller than
ours,   but   with the  same    spatial   resolution (16\arcmin).   In
\fig{fig_n288}, the  wavelet  decomposition clearly reveals  some wide
structures  missed by Grillmair  \etal\ (1995), especially towards the
south. (The arrows indicating the direction of the Galactic center in 
Grillmair \etal (1995) for NGC 288, NGC 362, and NGC 1904 are in error
- Grillmair, private communication). No dust emission from the IRAS 
100-\micron survey is detected.
NGC~288 is nearly free of observational biases, apart from some galaxy
clusters.  A  few such clusters of  galaxies are clearly detected (see
\fig{fig_n288}).  We suggest that the tidal radius determination could
be overestimated because of the presence of the clusters Abell~118 and
122, as already pointed out by Scholz \etal\ (1998) for Pal~5.

Tidal tails are  well separated in two  directions: first, towards the
galactic center (dashed arrow), second, aligned with  the orbit of the
GC (dotted  arrow).  NGC~288  has  recently undergone  a  gravitational
shock (Odenkirchen 1998).   It is very  likely   that the  tidal tails
visible in \fig{fig_n288}a are the results, in projection on the plane
sky, of both  the disk shocking and  the relics of  the bulge shocking
from the  last passage close    to the bulge  (Dauphole \etal\  1996).
NGC~288 exhibits  very important  tidal tails,  extending  up to
350~pc  from  the cluster:  this  has to  be    related to  its strong
interaction     with the    Galaxy,   as    found    by  GO97     (see
\tab{table_dynamic}).   We  count about 1200  stars  outside the tidal
radius of the cluster but  we did not  attempt an estimate of the mass
in the  outer  parts of the  cluster  because of the  poor photometry.
Further  CCD  observations with   deep and  precise  photometry should
provide very accurate mass loss rates.   From its orbital motion, this
cluster appears to be   a very good   candidate for tracing the  local
galactic potential (disk scale height and surface density).

\begin{figure*}[hpbt]
\caption{NGC~288.   ~~(a):   Surface  density  plot
displaying tidal tails (in Log)  around NGC~288.  The different arrows
indicate  the directions of the cluster  proper motion (dotted arrow),
of   the    galactic center  (dashed   arrow), and   of  the direction
perpendicular to the galactic plane (solid  arrow).  The dashed circle
centered  on the cluster  indicates  its tidal radius.  The horizontal
double  arrow   scale stands for     200~pc.   (b):
tidal-tail  density  overlaid  with  the surface  density  contours of
galaxies  ($>$3-$\sigma$)  at the  same  resolution.  
(c): Radial surface density profile with the power-law fit to our data
in the  external parts, while the  inner surface density profile comes
from the data (diamond) by Trager \etal\ (1995), shifted vertically to
fit our star count   data.   The vertical  arrow indicates   the tidal
radius.   (d):  Overdensities   of  galaxy  counts
overlaid with the positions of the Abell  clusters (triangle) known in
the same field.}
\label{fig_n288}
\end{figure*}

\subsection{NGC~1261}

NGC~1261 is a remote  cluster at a distance of  15.1~kpc from the sun,
with  its horizontal branch (HB) at  $V$ = 16.70~mag.   It has a tidal
radius  of  about  34~pc  and  a concentration  \conc\   =  1.27.  Its
evolution is probably driven by its internal dynamics (Zoccalli \etal\
1998, GO97).   Its  field is  not polluted  by  strong dust extinction
(E($B-V$) = 0.02), but the main bias is coming from the extra-galactic
object   overdensities.  Although no  Abell cluster  is present in the
field,  we detect the presence of  galaxy clusters  which are strongly
correlated  with  some    stellar     extensions  as     visible    in
\fig{fig_n1261}. We can  conclude here that the  N-E extension of  the
extra-tidal material, which is   aligned  with the direction of    the
galactic center (dashed arrow),  is  a real tidal  feature of  the GC,
because there is no strong galaxy cluster at this location.  The slope
\q{3}{6} (see \tab{slope_profile})  is probably highly contaminated by
back- and foreground stars and not useful.  Zoccali \etal\ (1998) find
evidence for mass  segregation in the  cluster ($x_{obs}=1.7  \pm 0.5,
r>4.4r_c$), segregation   which should   affect the tidal     tails as
discussed in Section~5.

\begin{figure*}[hbpt]
\caption{NGC~1261.  ~~(a): Surface density   plot
displaying tidal tails (in Log) around NGC~1261.  The different arrows
indicate  the direction of the  galactic center (dashed  arrow) and of
the direction perpendicular to the galactic  plane (solid arrow).  The
dashed circle centered on the cluster indicates its tidal radius.  The
horizontal double arrow  scale stands  for 100~pc.  
(b): Tidal-tail density overlaid  with the surface density contours of
galaxies   ($>$3-$\sigma$) at the  same  resolution.  
(c): Radial surface density profile with the power-law fit to our data
in the external parts,  while the inner  surface density profile comes
from the data (diamond) by  Trager \etal (1995), shifted vertically to
fit our  star  count data.   The  vertical  arrow  indicates the tidal
radius.  (d): Overdensities of galaxy counts; there
is no Abell galaxy cluster in this field.}
\label{fig_n1261}
\end{figure*}

\subsection{NGC~1851}

NGC~1851 is a  remote cluster at a distance  of 11.7~kpc from the sun,
with its horizontal  branch (HB) at $V$ =  16.15~mag.  It  has a tidal
radius of about  49~pc and a   very high concentration \conc\ =  2.24.
The western   part and   the S-W   part   of NGC 1851   extension  are
contaminated  by  galaxy clusters  (Abell~514 and anonymous)  and by a
bright  star   also   observed by   the  IRAS  100-\micron   map  (see
\fig{fig_n1851}).  The extinction  is not important  towards NGC~1851,
with E($B-V$)  = 0.02.  Stars   unbound  from the  cluster are  likely
tracing the   orbital  path, here these tails   seem to  have  a
preferential direction towards  the galactic center (dashed arrow  and
S-E  extension).   The   cluster position indicates  that   it  is not
suffering  a  strong     shock,    as  confirmed   by     the   ratio
$\nu_{tot}/\nu_{evap} =  1.0$  from  GO97, which   indicates  that the
evolution of this cluster  is mainly internally driven.  Consequently,
the surface  density profile  in the outer   parts  of the cluster  is
mainly  shaped by evaporation and tidal  stripping  at its location in
the Galaxy.  Saviane \etal\ (1998) found a  slight mass segregation in
this  cluster which affects  the tidal tail  detection by lowering the
mean mass of the unbound stars (Section~5).

\begin{figure*}[hbpt]
\caption{NGC~1851.   ~~(a): Surface density   plot
displaying tidal tails (in Log) around NGC~1851.  The different arrows
indicate  the direction of the  galactic center (dashed arrow), and of
the direction perpendicular to the  galactic plane (solid arrow).  The
dashed circle centered on the cluster indicates its tidal radius.  The
horizontal double arrow scale stands   for 100~pc.  
(b): Tidal-tail density overlaid  with the surface density contours of
galaxies ($>$3-$\sigma$)  at the  same resolution.  
(c): IRAS   100-\micron\  chart  overlaid with   the  above tidal-tail
surface  density contours.  (d): Overdensities of
galaxy counts overlaid with  the Abell clusters (triangle) detected in
the same  field. (e):  Radial surface density profile with
the power-law  fit to our data in  the external parts, while the inner
surface density profile comes from the data (diamond) by Trager \etal\
(1995),  shifted vertically to fit  our star count data.  The vertical
arrow indicates the tidal radius.}
\label{fig_n1851}
\end{figure*}

\subsection{NGC~1904 $\equiv$ M79}

NGC~1904  is a remote cluster  located at a  distance of 12.2~kpc from
the sun, with its horizontal branch (HB) at $V$ = 16.15~mag.  It has a
tidal radius of about  32~pc and a concentration \conc\  =  1.72.  NGC
1904 is surrounded  by a halo  of unbound stars (see \fig{fig_n1904}),
as previously seen by  Grillmair \etal\ (1995), on  a wider  field but
with a    lower spatial  resolution (them  with    16\arcmin,  us with
6.5\arcmin)  which blurred all  the small structures we observe around
the   cluster.  We do  not find  evidence   for a large southern tidal
extension  as   observed by  Grillmair     \etal\ (1995). The difference 
here could be accounted to the lower resolution used by them, one part 
of this large tail could be due to the southern galaxy clusters not well
separated. We point out that in their and our work we select stars below 
the completeness limit  ($\approx 19$ mag) , completeness fluctuation 
are another possibility to explain some differences, but not on such 
a large scale.
The  tail is oriented    in the  direction of the
galactic center (dashed arrow).  As in the  case of NGC~288, the tidal
radius determination may be  overestimated because of the  presence of
galaxy clusters close  to NGC~1904.  Nevertheless,  the tidal tails of
this cluster do not appear to be  correlated with the distribution of 
the
extra-galactic  objects.  The  dust  extinction  is  low  towards this
cluster  (E($B-V$) = 0.01) and the   fluctuations of the dust emission
are low as traced by  the IRAS 100-\micron map.   Because of the short
relaxation time of NGC~1904  ($t_{rh}= 8.8 \times  10^8$ yr), the mass
segregation should affect as well the stellar populations in the tidal
tails.  Since, following GO97, $\nu_{tot}$ is  about 30~\% higher than
$\nu_{evap}$, this may  indicate a slight  influence of the  galaxy on
this cluster.

\begin{figure*}[hbpt]
\caption{NGC~1904.    ~~(a):   Surface density   plot
displaying tidal tails (in Log) around NGC~1904.  The different arrows
indicate the direction  of the galactic  center (dashed arrow)  and of
the direction perpendicular to the  galactic plane (solid arrow).  The
dashed circle centered on the cluster indicates its tidal radius.  The
horizontal double  arrow scale  stands for 100~pc.   
(b): Tidal-tail density overlaid with  the surface density contours of
galaxies  ($>$3-$\sigma$)  at  the same  resolution.  
(c): Radial surface density profile with the power-law fit to our data
in the  external parts, while the  inner surface density profile comes
from the data (diamond) by Trager \etal\ (1995), shifted vertically to
fit our  star  count data.   The  vertical  arrow indicates  the tidal
radius.  (d): Overdensities  of galaxy counts; there
is no Abell galaxy cluster in this field.}
\label{fig_n1904}
\end{figure*}

\subsection{NGC~2298}

NGC~2298 is  a remote cluster located  at a distance  of 10.4~kpc from
the sun, with its horizontal branch (HB) at $V$ = 16.11~mag.  It has a
tidal radius of about 19~pc and a  concentration \conc\ = 1.40.  There
are background fluctuations owing to the dust along the line of sight,
as clearly traced by  the IRAS 100-\micron map (see  \fig{fig_n2298}).
We perform a  quite high tail/background~ S/N  CMD selection because of  the
high background density (see \fig{fig_cmd_tot}, but there is still a 
bias because of the dust
extinction, as seen in \fig{fig_n2298}.  There is a southern extension
towards  the galactic  center (dashed  arrow) which  is interrupted by
dust absorption.  Some parts of the  Eastern extension located at (x =
--50, y = --10) of the tidal tails may be questionable, because of the
stronger  dust presence, nevertheless the  lower absorption can hardly
explain  all  these overdensities, since  their  distribution does not
follow the minimum  IRAS   100-\micron  emission map.    Clearly,  the
overdensity at (x = 60, y = 60) is associated with a low IRAS emission
area.  Given   the  position and the  distance  of  NGC~2298 from  the
galactic center (15.1~kpc), the  southern extension is  likely tracing
its  orbital path  and  not  the result   of  gravitational shock,  as
indicated by the ratio $\nu_{tot}/\nu_{evap} = 1.0$ from GO97.  The low
value  \q{3}{6} = --0.25, likely  due  to the  small extensions in the
outer parts, is questionable.

\begin{figure}[hbpt]
\caption{NGC~2298.  ~~(a): Surface density plot displaying
tidal tails  (in Log) around NGC~2298.   The different arrows indicate
the direction of    the galactic center   (dashed arrow)  and  of  the
direction perpendicular to   the galactic  plane (solid arrow).    The
dashed circle centered on the cluster indicates its tidal radius.  The
horizontal double arrow  scale  stands for  100~pc.    (b):
IRAS  100-\micron\ chart  overlaid  with the  above tidal-tail surface
density contours.}
\label{fig_n2298}
\end{figure}

\subsection{NGC~4372}

NGC~4372 is a nearby globular cluster located at a distance of 4.6~kpc
from the sun, with its horizontal branch (HB) at  $V$ = 15.30~mag.  It
has a tidal radius of about  52~pc and a  concentration \conc\ = 1.30.
The   presentation of the detection  of  the overdensities around this
cluster illustrates the dramatic  influence of varying dust extinction
(see   \fig{fig_n4372}).   Strangely enough,  the very  elongated dust
filament observed in the IRAS 100-\micron\  map ends very close to the
cluster: this   may suggest  an  interaction of  the  cluster with the
interstellar  medium    currently  at   play.   Following  GO97   (see
\tab{table_dynamic}), this cluster has an evolution strongly driven by
the galaxy ($\nu_{tot}/\nu_{evap}=3.8$).

\begin{figure}[hbpt]
\caption{NGC~4372. ~~IRAS 100-\micron\ map  overlaid with the contours
of the overdensities in star-counts, which are completely disturbed by
the dust  extinction, particularly along  the elongated dust filament.
The different arrows  indicate the  direction  of the  galactic center
(dashed  arrow) and of  the   direction perpendicular to the  galactic
plane (solid  arrow).   The dashed   circle  centered on the   cluster
indicates its tidal radius. The  horizontal double arrow scale  stands
for 100~pc.}
\label{fig_n4372}
\end{figure}

\subsection{NGC~5139, $\omega$ Cen}

NGC~5139, the most  massive  galactic globular  cluster  (Meylan 1987,
Meylan \etal\ 1995, Merritt  \etal\ 1997), currently crossing the disk
plane, is  a nearby globular cluster located  at a distance of 5.0~kpc
from the sun, with its horizontal branch (HB) at $V$  = 14.53~mag.  It
has a tidal  radius of about 65~pc and  a concentration \conc\ = 1.24.
Its relative proximity allows to reach the main  sequence for the star
count selection.

Given the very good tail/background~ S/N ratio, we perform an absolute
calibration of   the photometry using the  data  from Cannon \& Stobie
(1973) and Alcaino \& Liller (1987) with an error which is still about
$\sigma$ = 0.2~mag.  Although obvious biases by dust absorption affect
the star counts, as seen,  e.g., at the positions (x  = 50, y =  --25)
and   (x  =    --50, y   =    --70)  on  the   IRAS   100-\micron  map
(\fig{fig_n5139}), there are two  large  and significant tidal  tails:
NGC~5139 is  releasing currently  some large  amounts of   stars.  The
tidal tail  extensions are perpendicular   to the galactic plane  (see
\fig{fig_n5139}), which is a clear  sign of disk-shocking, as observed
in our numerical simulations (CLM99).  By considering star-counts with
magnitude $B < 19$ (the completeness limit),  we found about 7000 $\pm
600$ stars outside one tidal  radius, in the $4^\circ \times 4^\circ$ field.  
This magnitude corresponds to a
0.63 \msun\  star  at a  distance of 5~kpc.    Assuming the same  mass
function in  the cluster and in the  tidal  extensions, because of its
large relaxation time,  we estimate a total  of 1.9 $10^4$ \msun\  for
the  escaped stars, with the assumption  of a Salpeter law ($\alpha$ =
--2.35) mass function for the stars down to 0.1  \msun. Thus the tidal
tails represent  about 0.6~\% of the  cluster  mass for  total cluster
mass of about 5.1 10$^6$ \msun.  This is consistent with the numerical
simulations (CLM99, Johnston \etal\ 1998) given the high uncertainties
on the mass function, the photometric calibration, the mass-luminosity
relation used  (see  e.g.,  Saviane \etal\   1998),  and the  possible
steeper mass  function in the tidal tails,  as discussed  in Section~5
for a slope $\alpha$= --2.8. We point out that a steeper mass function has
been observed in the halo of NGC~5139 (Anderson, 1998)

The \q{1}{2} parameter  value and the  position of the cluster  in the
galaxy  indicate  that  NGC~5139   is presently  experiencing   a disk
shocking,  with  an important  mass loss of   stars, whose presence is
clearly observed in  the  immediate neighborhood  of the  cluster. The
observed proper motion  of NGC~5139 indicates  that this cluster is in
the early phases of  its disk crossing.  This  confirms the high value
of the  ratio $\nu_{tot}/\nu_{evap} =  9.4$ estimated by GO97.  In the
case of NGC~5139, the disk-shocking consequences are combined with the
bulge-shocking ones, since the cluster  orbit goes as close as 1.8~kpc
from the galactic center (Dauphole \etal\ 1996).

We choose to present here  the same wavelet  planes that those for the
other clusters,  but  given the high density   --- significance --  of
NGC~5139   tidal  tails,  we  illustrate,  in  \fig{fig_resolution}, 
the different  spatial resolutions for  $\omega$ Cen after
filtering  of the background    noise.  It is  worth mentioning  that,
because of the   internal rotation of  this cluster   (Meylan \& Mayor
1986, Merritt, Meylan  \& Mayor  1997),  the global mass loss  rate is
enhanced by a  factor of  2  with respect  to  the N-body  simulations
(CLM99) and Fokker-Planck estimates (Longaretti  \& Lagoute 1996).  In
the  discussion we consider the effect  of the mass segregation on the
mass loss derivation.

\begin{figure*}[hbpt]
\caption{NGC~5139 $\equiv$ $\omega$~Centauri.  ~~(a):
Surface density plot displaying  tidal tails (in Log) around NGC~5139.
The different arrows  indicate  the directions of the  cluster  proper
motion (dotted arrow), of the  galactic center (dashed arrow), and  of
the direction perpendicular to the galactic  plane (solid arrow).  The
dashed circle centered on the  cluster indicates its tidal radius. The
horizontal   double arrow stands  for  100~pc.  (b):
IRAS   100-\micron\ chart overlaid  with  the above tidal-tail surface
density contours.    (c):  Radial  surface  density
profile with   the power-law fit  to  our data in the  external parts,
while the inner surface density profile comes  from the data (diamond)
by Trager   \etal\ (1995), shifted  vertically to  fit  our star count
data. The vertical arrow indicates the tidal radius.  }
\label{fig_n5139}
\end{figure*}

\subsection{NGC~5272 $\equiv$ M3}

NGC~5272 is a globular  cluster located at a  distance of 9.7~kpc from
the sun, with its horizontal branch (HB) at $V$ = 15.65~mag.  It has a
tidal radius of about 105~pc  and a concentration  \conc\ = 1.85.  The
cluster  is near the  edge of the  plate, preventing the  study of its
Eastern side (see \fig{fig_n5272}).   The field is  polluted only by 2
small galaxy clusters, viz.   Abell~1781  and Abell~1769,  the  former
being detected only at 2.5-$\sigma$ level.  Unfortunately, a defect on
the plate E131 (POSS) have blurred the extra-galactic object detection
(peak at x = 120, y = --20).  We emphasize that point-source detection
with SExtractor  is   less affected  by  this  defect.  There  is   no
anticorrelation at all between the tidal tails  and the dust emission,
as we checked with the IRAS 100-\micron  map, which is  at a low level
(E($B-V$) = 0.01).  The extension at (x = --30, y = --50), towards the
galactic  center (dashed arrow), is  the more reliable.  Thus from the
low value of the slope \q{1}{3} = --0.35,  we can infer that the field
pollution bias must be  quite strong, providing a rather constant radial
surface density.  The comparison  with the data from
Trager \etal\ (1995),  which  obtained star-count values smaller  than
our  data near  the  tidal  radius,   confirms this point.    The long
relaxation  time  of NGC~5272,  viz.     $t_{rh}=7.3 \times 10^9$  yr,
implies that the mass segregation  should not affect strongly the mass
function of the unbound stars.  Gunn \& Griffin (1979) found some weak
rotation  in this globular  cluster  which should slightly enhance the
mass  loss rate  by a  factor  1.1-1.2 (Longaretti  \&  Lagoute 1996).
There is no apparent correlation between the  tidal tail direction and
the proper motion of the cluster (dotted arrow).

\begin{figure*}[hbpt]
\caption{NGC~5272.    ~~ (a): Surface   density plot
displaying tidal tails (in Log) around NGC~5272.  The different arrows
indicate the directions of  the cluster proper motion (dotted  arrow),
of the   galactic  center   (dashed arrow),   and  of   the  direction
perpendicular to the galactic  plane (solid arrow).  The dashed circle
centered   on the cluster indicates  its  tidal radius. The horizontal
double  arrow scale stands for  200~pc.   (b): Above
tidal-tail   density  overlaid with the    surface density contours of
galaxies ($>$3-$\sigma$)  at  the same  resolution.  
(c): Radial surface density profile with the power-law fit to our data
in the external parts,  while the inner  surface density profile comes
from the data (diamond) by Trager \etal\ (1995), shifted vertically to
fit  our star count  data.   The  vertical  arrow indicates the  tidal
radius.    (d):    Overdensities of  galaxy  counts
overlaid with  the  Abell  clusters (triangle)  detected  in the  same
field. The extended structure is due to spurious detections because of
a defect on the plate (see text).}
\label{fig_n5272}
\end{figure*}

\subsection{NGC~5694}

NGC~5694 is a  very remote globular cluster located  at a  distance of
33~kpc  from  the  sun,  with its   horizontal branch  (HB)  at $V$  =
18.50~mag, which is  a strong limitation for  star counts.  Given its
large distance from  the galactic center, namely 27.5 kpc, this  
cluster is not expected  to suffer
strong gravitational  shocks ($\nu_{tot}/\nu_{evap}$ = 1.0,  GO97).   It
has a tidal radius  of about 41~pc and  a concentration \conc\ = 1.84.
We  select the   stars   on the giant  branch   only,  with  a  higher
tail/background~  S/N ratio in order to  avoid as much as possible the
galaxies   which are the      strongest  bias  in  this  field    (see
\fig{fig_n5694}).  The  lower  dust extinction,   mapped through  IRAS
100-\micron emission, could induce  an artificial extension in the S-W
part of the cluster, at the position (x = 20, y = --15).  But the huge
extension in the S-E  part can be  attributed to extra-tidal material,
with  high   confidence since   it  is correlated   with   higher dust
extinction and there is only one  small galaxy cluster at the position
(x = --20, y   = --3).  It must be   stars tidally stripped   from the
cluster, material which is now trailing/leading along the orbit of the
cluster.  As in the other clusters, it is aligned towards the galactic
center direction (dashed arrow), but it  might also be a projection effect
of its orbital plane with the galactic center  direction.  The size of
this extension is about  300~pc in the  sky and is probably even  much
larger  because  of the shallow  photometry  available on this distant
cluster.

\begin{figure*}
\caption{NGC~5694.    ~~(a):  Surface  density  plot
displaying tidal tails (in Log) around NGC~5694.  The different arrows
indicate the direction of the  galactic  center (dashed arrow) and  of
the direction perpendicular to the galactic  plane (solid arrow).  The
dashed circle centered on the cluster  indicates its tidal radius. The
horizontal double arrow  scale  stands for 200~pc.  
(b): Tidal-tail density overlaid  with the surface density contours of
galaxies ($>$3-$\sigma$)  at the   same resolution.  
(c):   IRAS 100-\micron\ chart     overlaid with the above  tidal-tail
surface density contours.    (d): Overdensities  of
galaxy counts; there is no Abell galaxy cluster in this field.}
\label{fig_n5694}
\end{figure*}

\subsection{NGC~5824}

NGC~5824 is a  very remote globular cluster  located at a  distance of
32.2~kpc  from  the sun, with   its horizontal  branch  (HB) at  $V$ =
18.60~mag, which  is a strong limitation for  star  counts.  At a large 
distance from the galactic center, namely 26 kpc,  this cluster is  
not expected 
to suffer strong gravitational shocks ($\nu_{tot}/\nu_{evap}$  = 1.6, GO97).   
It has a
tidal    radius of about 147~pc  and   a  concentration \conc\ = 2.45.
Because of  a  low tail/background~ S/N  ratio,  a consequence  of the
faint $V_{HB}$ magnitude, the overdensity  map around NGC~5824 appears
to  be very  noisy  (\fig{fig_n5824}).   Grillmair \etal\ (1995)  find
around this cluster   more extended structures,  aligned with  the N-S
direction, than we do in the same field: this may be partly due to our
rather shallow photographic films.   There are some strong biases  due
to dust   extinction  as it  can  be seen   on \fig{fig_n5824} at  the
position  (x = --20, y  = --20) and due  also  to some galaxies spread
mainly  over  the Southern  part.   GO97 indicate that NGC~5824 should
experience important  interactions  with the  tidal galactic field,  a
prediction  we  are not able     to confirm because   of the   tangled
observational biases.  But it appears that a preferential direction of
the  cluster extension could   be perpendicular to  the galactic plane
(solid arrow), either due to a disk shocking or tracing the orbital 
motion of the cluster. Nevertheless a bulge shocking effect cannot be 
ruled out in the case of a very eccentric orbit.

\begin{figure*}[hbpt]
\caption{NGC~5824.  ~~(a): Surface  density  plot
displaying tidal tails (in Log) around NGC~5824.  The different arrows
indicate the  direction of the  galactic center  (dashed arrow) and of
the direction perpendicular to the galactic  plane (solid arrow).  The
dashed circle centered on the  cluster indicates its tidal radius. The
horizontal  double arrow scale stands  for  400~pc. 
(b): Tidal-tail density overlaid with the  surface density contours of
galaxies  ($>$3-$\sigma$)  at  the  same resolution. 
(c): IRAS  100-\micron\ chart  overlaid    with the above   tidal-tail
surface  density  contours.   (d): Overdensities of
galaxy counts overlaid with the  Abell cluster (triangle) detected in
the same field.}
\label{fig_n5824}
\end{figure*}

\subsection{NGC~5904 $\equiv$ M5}

NGC~5904 is a globular cluster located at a distance of 7~kpc from the
sun, with its horizontal  branch (HB) at  $V$  = 15.06~mag.  It  has a
tidal  radius of about 66~pc  and a  concentration  \conc\ = 1.87.  We
present  only the S-E part  of  the tidal extensions (\fig{fig_n5904})
because of its position on the plate.  No bias due to dust is reported
towards this field (E($B-V$) = 0.03). The presence on a galaxy cluster
close to the  tidal radius of the   cluster enhances artificially  and
locally the tidal  tail pointing towards  the galactic  center (dashed
arrow) and the  direction perpendicular to  the galactic plane  (solid
arrow).  Nevertheless, it  is  obvious that  an extension  is  present
towards this direction, since the galaxy cluster size is significantly
smaller than the  size of the  globular cluster extension, as shown on
\fig{fig_n5904}b with the same resolution used for the star and galaxy
surface densities.  Lehmann \&  Scholz (1996) found already indication
of tidal tail around this  cluster from its surface brightness profile
which departs from  a King profile; this may  be explained as  well by
the galaxy cluster near  the  tidal radius.  From Odenkirchen   \etal\
(1997), NGC~5904 is  just beginning its crossing  through the disk and
towards the galactic center.  Consequently, we could observe the first
effect of the  gravitational shocking on this  cluster, with the  tail
aligned towards the tidal directions (see CLM99) after being
compressed during the crossing. Indeed the momentum transfer to the
cluster stars is in the Z direction during the disk shocking. From
GO97, NGC~5904
suffers  strong        interactions    with  the    galaxy,       with
$\nu_{tot}/\nu_{evap}$  = 26, a   high  value due  to  the use  of the
Bahcall \etal\ (1983) galactic model which enhances the gravitational
shocks because of its nuclear component and the form of the disk
potential which does not vanishes at the center as it is the case for
the model from Ostriker \& Caldwell (1983).

\begin{figure*}[hbpt]
\caption{NGC~5904  $\equiv$  M5.    ~~ (a):  Surface
density  plot  displaying tidal tails  (in  Log) around NGC~5904.  The
different arrows indicate the directions  of the cluster proper motion
(dotted  arrow), of the galactic  center  (dashed arrow),  and of  the
direction  perpendicular  to the  galactic plane  (solid  arrow).  The
dashed circle centered on the cluster indicates its tidal radius.  The
horizontal double  arrow scale stands  for 100~pc.  
(b): Tidal-tail density overlaid with the  surface density contours of
galaxies ($>$3-$\sigma$) at  the  same resolution.  (c):
Overdensities  of  galaxy counts overlaid    with  the Abell  cluster
(triangle) detected in the same field.}
\label{fig_n5904}
\end{figure*}

\subsection{NGC~6205 $\equiv$ M13}

NGC~6205 is a  globular cluster located at  a distance of 6.8~kpc from
the sun, with its horizontal branch (HB) at $V$ = 14.90~mag.  It has a
tidal radius  of about 56~pc and a  concentration  \conc\ = 1.49.  The
bias towards   NGC~6205  are not strong    as shown by   the weak IRAS
100-\micron flux and the relatively high tail/background~ S/N ratio in
the  CMD. Given the  position of the cluster  on the survey plates,
we extract a field of 90\arcmin\ in size.  There  is no strong bulk of
tidal stars  (see \fig{fig_n6205}) due  to any shock  and the field is
too small  to  detect any  large scale  structure corresponding to the
orbital path.  An extension  can be seen  towards the galactic  center
(dashed arrow) at the position (x =  --10, y = --25), although located
inside   the tidal  radius,  which  highlights  the  limitation  of an
azimuthally  averaged radial surface  density.  This  extension is not
correlated with the proper motion. We note that an extended default on 
the plate center worsens the cluster/background star separation.

\begin{figure}[hbpt]
\caption{NGC6205.  ~~ Surface  density plot displaying tidal tails (in
Log) around NGC~6205.  The different arrows indicate the directions of
the cluster proper   motion  (dotted arrow),  of  the galactic  center
(dashed  arrow), and of  the direction  perpendicular  to the galactic
plane   (solid arrow).   The dashed  circle   centered  on the cluster
indicates its  tidal radius. The  horizontal double arrow scale stands
for 100~pc.}
\label{fig_n6205}
\end{figure}

\subsection{NGC~6254 $\equiv$ M10}

NGC~6254 is a nearby globular cluster located at a distance of 4.1~kpc
from the sun, with its horizontal branch (HB) at $V$  = 14.65~mag.  It
has a tidal  radius of about 26~pc and  a concentration \conc\ = 1.40.
This  cluster is a  striking case because  of a strong gradient in the
dust extinction as seen on  \fig{fig_n6254} with the IRAS  100-\micron
map.  The  southern extension anticorrelates quite  well with the dust
emission which is the sign of a possible bias.  An obvious decrease of
the stellar surface density  is correlated with  the dust emission  at
the position  (x = --40, y =  --30).  Nevertheless the inner NE-SW and
the northern extensions are not anticorrelated with the dust emission.
The second break  at  $\log(r)\simeq 1.6$, apart from the one around
the tidal radius,  in the  radial  density profile (see 
\fig{fig_n6254}c) must correspond  to a very recent disk-shocking, with
the  diffusing stars  (cf.  CLM99) still  close to  the cluster.  This
conclusion is  strengthenedd by the   proper motion of  the cluster whose
direction  (dotted arrow) is opposite   to  the disk direction  (solid
arrow): Odenkirchen \etal\ (1998 ) indicate  that NGC~6254 suffered its
last   disk crossing  about  20~Myr   ago.  Considering  the  northern
extension as a genuine tidal tail made of stars  from NGC~6254, we can
give  a lower limit  for the diffusion  velocity, which is a projected
expansion  velocity  of the tidal   material in the  cluster reference
frame: at  a distance of 4.1~kpc, for  a projected distance of 150~pc,
we  obtain about 7~\kms\ as a  lower limit of  the diffusion velocity.
We note that the velocity dispersion  of stars in NGC~6254 is similar,
with $\sigma_0$ = 6.6  \kms\ (Pryor  \&  Meylan 1993).  This  velocity
diffusion probes the  differential diffusion of  stars released in the
Galaxy along with  the global dynamical friction  of the cluster which
is not  felt by the  unbound stars. Actually this diffusion velocity
is surprisingly high compared to the dispersion velocity, where we
would expect low velocity dispersion for the unbound stars: a 
misclassification of these clumps as genuine cluster stars or an
underestimation of the last crossing time cannot be ruled out.
Given  the quite short relaxation
time $t_{rh} = 7.6 \times 10^8$ yr, the mass segregation must be present
in this cluster,  even though Hurley  \etal\  (1989) did not find   any
evidence.  Such a mass segregation must  lead to a steep mass function
in the tidal tails.

\begin{figure*}[hbpt]
\caption{NGC~6254.   ~~(a): Surface   density  plot
displaying tidal tails (in Log) around NGC~6254.  The different arrows
indicate  the directions of the  cluster proper motion (dotted arrow),
of the   galactic center  (dashed    arrow),  and of   the   direction
perpendicular to the galactic plane (solid  arrow).  The dashed circle
centered on the cluster   indicates its tidal radius.  The  horizontal
double  arrow  scale stands for 100~pc.   (b): IRAS
100-\micron\ chart overlaid with  the above tidal-tail surface density
contours.  Lower-right panel (c):  Radial surface density profile with
the power-law fit  to our data in  the external parts, while the inner
surface density profile comes from the data (diamond) by Trager \etal\
(1995),  shifted vertically to fit our  star count data.  The vertical
arrow indicates the tidal radius.}
\label{fig_n6254}
\end{figure*}

\subsection{NGC~6397}

NGC~6397 is  a very nearby globular  cluster located  at a distance of
2.2~kpc   from the sun,  with  its  horizontal  branch (HB)  at  $V$ =
12.87~mag.  It  has a tidal radius of  about 66~pc  and a concentration
\conc\  =  2.50.  It is  the only  post core-collapsed cluster  in our
sample,  although   NGC~1851  and  NGC~5824 have   also   rather large
concentrations.   This   is  the second   example,   with NGC~4372, of
overdensities strongly biased by dust extinction fluctuation as it can
be  seen in \fig{fig_n6397}.   All   the overdensities  found  in  the
northern   and  eastern   parts cannot    be disentangled  from   dust
extinction. Only the S-E extension,  at the position  (x = --100, y  =
--100),   could be a    genuine tidal tail,  but  with  a somewhat low
confidence in spite of the fact that these star counts are more than 3
$\sigma$ above the background because the dust extinction fluctuations
in  this   field    are  quite    high  ($\sigma^2 (S_{100})    \simeq
\overline{S_{100}}  \simeq 20$  MJy/sr for  the  IRAS-100\micron flux).
Nevertheless, we emphasize that this extension is perpendicular to the
plane (solid arrow) as expected  for disk  shocking (CLM99), thanks to
the momentum transfer in the Z direction. During the disk crossing the 
gained acceleration for the cluster stars is directed towards the 
cluster equatorial plane  parallel to the galactic plane. Then
the energy gained is released in this direction, perpendicular to the
galactic plane. The  mass
segregation found by Mould  \etal\ (1996) in  this cluster will affect
the mass function of the tidal tails. A  weak rotation of NGC~6397 has
been found (Meylan \& Mayor  1991) which should  enhance the mass loss
rate by about 20~\%, using Fig.~7 of Longaretti \& Lagoute (1996).

\begin{figure}[hbpt]
\caption{NGC~6397.  ~~(a): Surface density plot displaying
tidal tails (in Log)  around NGC~6397.  The different  arrows indicate
the  directions of the cluster  proper  motion (dotted arrow), of  the
galactic center (dashed arrow),  and of the direction perpendicular to
the galactic  plane (solid arrow).  The dashed  circle centered on the
cluster indicates its tidal radius.  The horizontal double arrow scale
stands  for 50 pc.  (b): IRAS 100-\micron\ chart overlaid
with the above tidal-tail surface density contours.}
\label{fig_n6397}
\end{figure}


\subsection{NGC~6535}
 
NGC~6535 is  a globular cluster located  at a distance of 6.6~kpc from
the sun, with its horizontal branch (HB) at $V$ = 15.73~mag.  It has a
tidal  radius of about 17~pc  and a concentration  \conc\  = 1.30.  In
spite of the high tail/background~  S/N color selection  on the CMD in
order to    avoid the high   background  density, the  cluster density
remains lower  than  the background  (see \fig{fig_n6535}).  The  dust
extinction induces a strong bias towards  this field (E($B-V$) = 0.32)
as seen in the IRAS 100-\micron map on \fig{fig_n6535}.
Part of the  northern extension could be  artificially enhanced by the
local lower dust extinction.  It is likely that the Southern extension
is lowered by local higher dust extinction at the position (x = 0, y =
--20).   As   indicated by GO97,   the  evolution of  this  cluster is
influenced by  the galactic potential ($\nu_{tot}/\nu_{evap}$  = 1.4).
Currently,  NGC~6535 is experiencing a strong  bulge shocking and disk
shocking  as indicated   by  the  correlation  of  the tail   with the
disk/bulge direction  (solid  and  dashed arrows,   respectively)  and
confirmed by its location in the galaxy,  viz. 1.2~kpc above the plane
and 4~kpc from the galactic center (Harris 1996).

\begin{figure*}[hbpt]
\caption{NGC~6535.  ~~  (a): Surface  density   plot
displaying tidal tails (in Log) around NGC~6535.  The different arrows
indicate the direction of the  galactic center  (dashed arrow) and  of
the direction perpendicular to the galactic  plane (solid arrow).  The
dashed circle centered on the cluster indicates its tidal radius.  The
horizontal  double arrow  scale stands  for  50 pc.  
(b): IRAS  100-\micron\  chart overlaid   with  the above   tidal-tail
surface density   contours.   (c):    Radial surface
density profile with  the power-law  fit to our  data  in the external
parts, while the inner  surface density profile   comes from the  data
(diamond) by Trager \etal\ (1995),  shifted vertically to fit our star
count data.  The vertical arrow indicates the tidal radius.}
\label{fig_n6535}
\end{figure*}

\subsection{NGC~6809 $\equiv$ M55}

NGC~6809 is a  globular cluster located at a  distance of 5.1~kpc from
the sun, with its horizontal branch (HB) at $V$ = 14.40~mag.  It has a
tidal radius of  about 23~pc and a concentration  \conc\ =  0.76.  The
overdensities (see \fig{fig_n6809})  are  strongly anticorrelated with
the dust  emission  traced by the  IRAS 100-\micron  map, e.g.  at the
position (x =  --50, y = --50), where  the dust extinction is probably
disturbing the tail surface density.  It may also be possible that the
Western extension (x = 90, y = 0) towards the galactic center (dashed
arrow) could   be associated  with   the cluster,  because it   is not
anticorrelated with the dust emission; the same remark applies for the
extension at (x = --30, y = --30).  The study of such a cluster should
greatly benefit from  the better transparency  to  the dust absorption
offered in $J$ and $K$ bands. In a previous study in V and 
I bands, Zaggia et al. (1997) found already evidence for cluster
stars in the halo of this object.

\begin{figure*}[hbpt]
\caption{NGC~6809  $\equiv$  M55.   ~~(a):  Surface
density plot  displaying tidal  tails (in Log)  around  NGC~6809.  The
different arrows indicate the direction of the galactic center (dashed
arrow) and of the direction perpendicular to the galactic plane (solid
arrow).  The dashed circle centered on the cluster indicates its tidal
radius.  The horizontal   double    arrow scale stands  for    100~pc.
(b): IRAS 100-\micron\ chart overlaid with the above
tidal-tail  surface density  contours.  (c): Radial
surface density profile  with the  power-law  fit to our  data in  the
external parts, while the inner surface density profile comes from the
data (diamond) by Trager \etal\ (1995), shifted  vertically to fit our
star count data.  The vertical arrow indicates the tidal radius.}
\label{fig_n6809}
\end{figure*}

\subsection{NGC~7492}

NGC~7492 is  a  remote globular    cluster located at  a  distance  of
24.3~kpc from  the sun,  with  its horizontal  branch  (HB) at   $V$ =
17.63~mag.  It has  a tidal radius of about  62~pc and a concentration
\conc\ =  1.0.  There is  no dust emission towards  this field and the
background galaxy clusters   are  located far  from the    cluster, as
indicated  in \fig{fig_n7492}.   Obviously,   the  overdensity  at the
position (x  =  18, y = 25)    is associated with  the galaxy  cluster
Abell~2533.  Because of the low mass of this cluster, GO97 find a fast
evolution  in  the Galaxy field,   with $\nu_{tot}/\nu_{evap} = 77.8$,
compared to  its intrinsic evolution.    Clearly, a tiny  extension is
visible,  pointing towards the galactic  center  (dashed arrow).  This
lack of tidal  extension is not  in contradiction  with the conclusion
drawn by GO97, given its current  location far from  the center of the
Galaxy (23.5~kpc).   A  higher tail/background~  S/N ratio  selection,
using  high-quality CCD  data, may  allow  the  detection of  very low
surface density extension   related to tidal  tails extending  away  from
NGC~7492.

\begin{figure*}[hbpt]
\caption{NGC7492.   ~~ (a):  Surface  density  plot
displaying tidal tails (in Log) around NGC~7492.  The different arrows
indicate the direction of the   galactic center (dashed arrow) and  of
the direction perpendicular to the galactic  plane (solid arrow).  The
dashed circle centered on the cluster indicates its tidal radius.  The
horizontal double arrow  scale stands  for 100~pc.  
(b): Tidal-tail density overlaid with the  surface density contours of
galaxies ($>$3-$\sigma$)  at  the same resolution.   (c):
Overdensities of   galaxy  counts  overlaid  with  the Abell   cluster
(triangle) detected in the same field.}
\label{fig_n7492}
\end{figure*}

\subsection{Palomar~5}

Palomar~5  is a remote   globular  cluster located   at a  distance of
21.8~kpc  from  the sun,  with its  horizontal  branch (HB)   at $V$ =
17.63~mag.  It has a tidal radius of  about 107~pc and a concentration
\conc\ = 0.74.  It  is one of the  most  remote cluster with  measured
proper motions (Schweitzer \etal\ 1993, Scholz \etal\ 1998). The tidal
radius  could be lower than  previously measured, down to 7\arcmin, in
agreement with  its orbit (Scholz  \etal\ 1998).  In  \fig{fig_pal5} we
present the overdensities, which are strongly biased by the background
galaxy clusters present in the field. Because of the unreliable 
star/galaxy separation above $R\approx 18$, the confusion is quite
strong for this remote and faint cluster.
As pointed out already by Scholz
\etal\ (1998), the galaxy cluster  Abell~2050 could be responsible for
the previous (commonly adopted) overestimate of the tidal radius.  The
dust extinction is very  weak in this field  (E($B-V$) = 0.03), and do
not exhibit any anticorrelation with the  overdensities, as checked on the
IRAS   100-\micron map.  The  background  galaxy  distribution and the
large  distance to this cluster make  difficult any  conclusion on the
genuine location, if any, of stars stripped from the cluster.

\begin{figure*}[hbpt]
\caption{Palomar~5.    ~~ (a):  Surface  density plot
displaying tidal tails (in   Log) around Pal~5.  The  different arrows
indicate  the directions of  the cluster proper motion (dotted arrow),
of   the    galactic center   (dashed  arrow),   and of  the direction
perpendicular to the galactic plane  (solid arrow). Here, the direction of
the galactic center is similar, in projection, to the direction perpendicular
to the galactic plane. The dashed
circle  centered on the    cluster  indicates its tidal   radius.  The
horizontal double arrow scale  stands for  200 pc.  
(b): Tidal-tail density overlaid with  the surface density contours of
galaxies   (3-$\sigma$) at the    same  resolution.  (c):
Overdensities of  galaxy   counts overlaid  with   the Abell  clusters
(triangle) detected in the same field.}
\label{fig_pal5}
\end{figure*}

\subsection{Palomar~12}
\begin{figure*}[hbpt]
\caption{Palomar~12.   ~~(a): Surface density  plot
displaying tidal tails (in  Log) around Pal~12.  The different  arrows
indicate the direction of  the galactic center  (dashed arrow)  and of
the direction perpendicular to the  galactic plane (solid arrow).  The
dashed circle centered on the cluster indicates its tidal radius.  The
horizontal  double arrow scale stands  for 200  pc.  
(b): Tidal-tail density overlaid with  the surface density contours of
galaxies ($>$3-$\sigma$)  at the same   resolution.  (c):
Overdensities  of galaxy counts overlaid  with   the Abell  clusters
(triangle) detected in the same field }
\label{fig_pal12}
\end{figure*}
Palomar~12 is   a remote globular  cluster  located  at a  distance of
17.8~kpc from  the sun,  with its  horizontal  branch  (HB) at   $V$ =
17.13~mag.  It  has a tidal radius of  about 49~pc and a concentration
\conc\ = 0.90.  The dust extinction is very low (E($B-V$) = 0.02), but
the contamination by background galaxy clusters is very important (see
\fig{fig_pal12}), although only two Abell galaxy clusters are reported in this
field.    The N-S oriented   very long  tail  is  contaminated by some
galaxies as  shown at position (x  = 15, y   = 30) in \fig{fig_pal12}.
Nevertheless this  tail is  a genuine  feature  made of stars  tidally
stripped, as shown by  the distribution of  the  galaxies as the  same
resolution.  The western  and eastern overdensities are related mainly
to  galaxies.  A higher   tail/background~ S/N CMD selection confirmed
the Pal~12 membership  of the top and bottom  clumps at positions (x =
0, y = $\pm$60). To get  an estimate of the  time of the last, if any,
gravitational  shock on this cluster,  we assume that these two latter
star clumps  are  remains of the   last shock.   Adopting  a diffusion
velocity for the tidal stars equal  to 1~\kms, similar to the velocity
dispersion (Djorgovski \& Meylan 1994) of such  a low mass cluster ($2
\times 10^4  \msun$) and assuming  the distance in  projection between
the clumps and the  cluster to be 350~pc,  we estimate 350~Myr as  the
time  since the  last shock.  This  is  a lower  limit  because of the
projection effect and the limited field.  Contrary to most other tidal
tail directions, the extension is perpendicular to the galactic center
direction  (dashed arrow) and  is in a plane  parallel to the galactic
disk.


\section{Discussion}
 
\begin{figure}[hbpt]
\caption{ The slope values \q{1}{3}
(between $\rt$ and 3$\rt$) versus \q{3}{6} 
(between 3$\rt$ and 6$\rt$) in  clusters with  wide
enough field. The dotted line stands for \q{1}{3} = \q{3}{6}.}
\label{fig_slope_gc}
\end{figure}
The  detection  of  stars  tidally   stripped from  globular  clusters
emphasizes  strongly the importance    of  the interactions of   these
stellar systems with the  Galaxy. In the  light of the possible biases
present  in  the above  observed  fields, it  is   possible to give an
estimate of the   physical status  of  the clusters   relative to  the
gravitational   shocks  they    suffer   in  the Galaxy     (see Table
\ref{tail_summary}).  In   the   case    of a  cluster    experiencing
disk-shocking  only,  it  will be first  compressed  in the  direction
perpendicular to the galactic plane, during the short time of the crossing;
then the  tidally  released  stars form  tails  perpendicular  to  the
galactic plane  (see,  e.g., NGC~5139).    In the  case of a   cluster
experiencing bulge-shocking  only, i.e., not too   far from the Galaxy
center,  the tails are   elongated mainly along  the  galactic density
gradient (spherical symmetry) and one can expect a correlation between
the tidal tail  direction and the galactic center.   This is true also
for   the more  general  case   of  galaxy-shocking, when  bulge-  and
disk-shocking are both at play, i.e., when the cluster is close to the
galactic center.  If the cluster has not experienced for a long time a
gravitational shock, its tidal tails   are on a large scale  oriented
along its orbit. However Grillmair (1992) showed using N-body simulations,
without any disk potential, that  strong ``bars'' orthogonal to the
orbital path will develop naturally near the apogalactica of the
cluster's orbit. 
In Table \ref{tail_summary} we, tentatively, give
the processes at play for creating the recent mass loss in the
clusters: it is based on tidal tails shapes, but, as well, on their
positions in the Galaxy and their orbit and proper motions, when they 
are available. It  explains the discrepancy between some tidal 
tail orientation and the type of physical process. We point out that
the projection effect must be important in some case (e.g NGC~288).
It has to be noted that a combination of disk- and bulge-shocking are 
expected to confuse the above simplified scenario (e.g. NGC~6535).

The case of NGC~5904 is interesting  since its proper motion is known:
the small tidal extension   observed is perfectly aligned towards  the
galactic center and the direction  perpendicular to the galactic plane
and not with its  motion along its orbit.  Given  its position  in the
Galaxy, this  cluster is  probably suffering   a  weak disk and  bulge
shocking.

\begin{figure}[hbpt]
\caption{Evolution in N-body  simulations of the surface density slope
for a power law between a  radius of 30 and  40 pc (solid line) and 40
and 50 pc (dash-dotted  line) for  a  globular cluster on  polar orbit
(CLM99).  The  disk crossing occur  at t =  10, 70, 150, 240, 310, and
380 Myr.  The crossing at t = 10, 150, 240 and 380 are clearly visible
on  the     variations    of   $\alpha(30\mbox{--}40\mbox{pc})$    and
$\alpha(40\mbox{--50}\mbox{pc})$.}
\label{fig_slope_simul}
\end{figure}

In  \fig{fig_slope_gc} we  show  the   slope  values \q{1}{3}   
(between $\rt$ and 3$\rt$) versus \q{3}{6} 
(between 3$\rt$ and 6$\rt$) 
for  the few clusters where it  is possible  to measure these
two parameters.  We  emphasize that  these  slope values are  probably
overestimated, especially for \q{1}{3},  because of central  crowding.
As  found  in dynamical simulations by   Johnston \etal\ (1998) and in
other observations  by  Grillmair \etal\  (1998) on   different radius
ranges, the mean  slope value for  \q{1}{6} is $-0.91  \pm 0.24$.  The
coefficient \q{3}{6} presents a  strong scatter (1.69) around its mean
value equal to   $-1.0$.  Its determination  is  difficult because the
stars no  more bound to  the cluster have  a  very low density  in the
outer  parts where the noise    dominates (see, e.g., NGC~2298).   The
quantity \q{1}{3} must be  a reliable indicator  of the recent  mass
loss  from the cluster, with  a steep slope  for the cluster suffering
shocks.   Then  the diffusion  of the  heated  stars  will flatten the
surface density profile.

In  \fig{fig_slope_simul}, we  present, from  our  N-body simulations
(CLM99),  the variation with time of  the surface density slope fitted
on a power law for two different ranges of radii.  It is remarkable to
note  the strong variation  of the slopes  during  the crossing of the
galactic plane.  Moreover  there is a  delay between  the variation of
the           $\alpha(30\mbox{--}40\mbox{pc})$     slope           and
$\alpha(40\mbox{--50}\mbox{pc})$ slope.  Here  the  dumping frequency of
the simulations  is  too low  to allow any   estimate of the diffusion
velocity of the bulk of stars stripped during the crossing. It appears
nevertheless to be lower than the velocity dispersion of the simulated
globular cluster ($\simeq$ 8 \kms).

We may link the case of NGC~6254  (see \fig{fig_n6254}) to the surface
density  profile computed from  our N-body  simulations (CLM99) before
and after  the  crossing  and  displayed  in  \fig{fig_profil_simul}).
Clearly    the  second break, at a radius $r>r_t$, in    the   observed  
cluster   density profile
($\log(r)\simeq1.6$)   and    simulated   cluster     density  profile
($\log(r)\simeq1.9$) indicates that disk shocking is currently at play
on   NGC~6254  and the   halo of  unbound  stars  has not yet diffused
outwards. Even if other mechanisms could produce such break (e.g. ``bars''
at the apogalactica radius) we note that this NGC~6254 is currently
just 1.6 kpc above the galactic plane.

The variations of  the \q{1}{3}  coefficient  between clusters with  a
strong  galaxy-driven  evolution are  expected    to be  important  as
observed in the simulations. Nevertheless this coefficient is, as
well, dependent on the orbital phase as shown by Grillmair (1992).

From our  N-body simulations  (CLM99),  using  multi-mass  King-Michie
models,  we  show that  the tidal  tails  are populated  mainly by the
lighter   stars of the  pruned  globular cluster,  because of its mass
segregation. In \fig{fig_slope_mf_m1}, we present the evolution of the
mass function slope (assumed to be a  power-law) for a simulation with
a globular  cluster  on polar orbit (CLM99).    The  duration of  this
simulation is too short to observe strong changes in the mass spectrum
through  the cluster  itself,  nevertheless it   can be  seen that the
radius of constant  mass function slope  is slightly  expanding during
the 800 Myr of  the simulation.  This  is especially true in the inner
parts of the cluster (see e.g. the isocontour $\alpha=-2$ on 
\fig{fig_slope_mf_m1}).

\begin{table*}
\begin{tabular}{lcccccccc}
\hline
\hline
Cluster & Observed size & Bias & Reliability & R$_\odot$& R$_{GC}$ 
& Z & Alignment & Type$^\dagger$ \\
Name    &  (pc)             &       &       & (kpc) &      (kpc)       &  (kpc)   
   &         \\
\hline
NGC 104 &  150          & SMC   & 4 & 4.1       & 7.3 &  -2.9 & 1/2 & DS/BS \\
NGC 288 &  350          & gal.  & 1 & 8.1  & 11.4 & -8.1 & 1/2      & DS/BS\\
NGC 1261 & 100          & gal.  & 3 & 15.2 & 17.1 & -12. & 2        & BS/OP\\
NGC 1851 & 130          & dust+gal. & 2& 11.7 & 16.3& -6.7 & 2      & OP\\
NGC 1904 & 130          & gal.  & 1 & 12.2 & 18.1& -6. & 1/2        & DS/OP\\
NGC 2298 & 200          & dust  & 3 & 10.4 & 15.4& -2.9 & 2         & OP/BS\\
NGC 4372 &              & dust  & 5 & 4.6 & 6.9 & -0.8 & --         & --\\
NGC 5139 & 170          & dust  & 2& 4.9 & 6.3& 1.3 & 1             & DS\\
NGC 5272 & 150          & gal.  & 4  & 9.7 & 11.6& 9.5 & 1          & DS\\
NGC 5694 & 300          & gal.  & 1 & 33. & 27.5& 16.7 & 2          & BS/OP\\
NGC 5824 & 450          & dust+gal. & 4 & 32.2 & 26.& 12.1 & 1      & DS/OP\\
NGC 5904 & 80           & gal.  & 2 & 7. & 6.& 5.1 & 1/2            & DS/BS\\  
NGC 6205 & 60           &       & 1 & 6.8 & 8.2& 4.4 & 2            & BS\\
NGC 6254 & 150          & dust  & 4 & 4.1 & 4.7& 1.6 & 3            & OP \\
NGC 6397 &              & dust  & 5 & 2.2 & 6.1& -0.5 & --          & --\\
NGC 6535 & 60           & dust  & 3 & 6.6 & 3.9& 1.2 & 2            & DS+BS \\
NGC 6809 & 190          & dust  & 4 & 5.1 & 4.& -2. & 3             & OP\\
NGC 7492 & 100          & gal.  & 1 & 24.3 & 23.5& -21.8 & 2        & BS\\
Pal 5   & 200           & gal.  & 4 & 21.8 & 17.2& 15.7 & 3         & OP\\
Pal 12  & 400           & gal.  & 3 & 17.8 & 14.7& -13.2 & 3        & OP\\
\hline
\end{tabular}
\caption{ Characteristics  of the tails.   We  indicate a reliability
level from 0  (reliable, no observational bias)  to 5 (unreliable) for
the observed  overdensities (tidal tails)  around these clusters.  The
position of the  cluster in the  galaxy is given through its distance
to the sun  ($R_\odot$), to the galactic  center (R$_{GC}$) and to the
plane (Z).  We give an indication of  the alignment of the tidal tails
perpendicular to  the galactic plane  (1),  aligned with  the galactic
center (2)  and  with no correlation  relative  to  any of   these two
directions (3).
~$^\dagger$  OP: Orbital path, DS:  disk shocking, BS: bulge shocking
(+ means both processes  are probably at  play; / means probably one
of the two processes).}
\label{tail_summary}
\end{table*}

Let us  compare the amount of  tidally stripped stars  obtained in our
simulations and observations.   Because of the magnitude limitation of
the plates and  films, it is likely that  we underestimate the observed
tidal tails.  The mass of the tidal tails  in the case of NGC~5139 has
been computed for a Salpeter law: if we assume a steep slope $\alpha$=
--2.8 for  the mass function  in the tidal tails,  we get  a tail mass
equal to  about 1~\% of the  total  mass of  the  cluster, equal to 5.1
10$^6$ \msun.  In spite of the great uncertainty on the star counts of
the tidal  tails, such a large  mass is an  upper limit  for the tidal
tail  mass from  the   simulations (CLM99).    It confirms both   that
NGC~5139 has a  genuinely large  total  mass (Meylan \etal\  1995) and
that the spectrum mass is likely less steep  than $\alpha=-2.8$ in the
outer part.   All these considerations  point  toward  a mass for  the
tidal tails between 0.6 and 1.0 \% of the total mass of NGC~5139.
From N-body simulation performed by Moore (1996) we can note that the
presence of tidal tails  is the indication of low dark matter content
in globular clusters.

NGC~7492 and Pal~12 provide an interesting comparison because of their
similar characteristics:  low masses (6  and  2 $\times$ $10^4$ \msun,
respectively), low concentrations  (1.0  and 0.9, respectively),   and
large distances    from  the  galactic center   (23.5   and  14.7 kpc,
respectively).  They are also  both strongly influenced by the Galaxy:
GO97 compute   $\nu_{tot}/\nu_{evap}$ = 77.8  and 17.9,  respectively.
Nevertheless, their tidal tails appear strongly different, with a very
extended structure for   Pal~12 and very  tiny one  for NGC~7492.  The
last gravitational  shock suffered  by  NGC~7492, if any,  has to have
occured much before the last one for  Pal~12.  The long time since the
last  gravitational shock suffered  NGC~7492  has allowed the  surface
density  of  the unbound stars to  fade  along the cluster  orbit.  We
estimate the last tidal shock suffered by Pal 12 to be about 350 Myr.

\begin{figure}[hbpt]
\caption{Surface  density  profiles  (Log)  from   N-body  simulations
(CLM99)   at t=230 Myr   (solid    line) and t=275  Myr   (dash-dotted
line). The disk crossing occurs at 240 Myr.}
\label{fig_profil_simul}
\end{figure}

\section{Conclusions}

We  have  observed 20 galactic   globular  clusters  with  multi-color
Schmidt plates  and films on wide  fields. Field and cluster stars are
sorted in the   color-magnitude    plane. A star-count analysis     is
performed on the color selected stars  to study the overdensities that
can be attributed to the stars stripped  from the globular clusters by
tidal shocks   (disk/bulge)  as   well  as  from   internal  dynamical
evolution.  We use the wavelet transform in order to enhance the weak 
tidal structures  at  large scales and   in  order to  filter  the high
background  noise at low galactic  latitude.  After highlighting  the
observational biases   resulting from  dust extinction  and background
galaxy clustering   at  faint  magnitudes,  we  reach   the  following
conclusions:
\begin{itemize}

\item All   the clusters  observed,  which do  not suffer  from strong
observational  biases,  present  tidal tails, tracing  their dynamical
evolution in the Galaxy (evaporation,  tidal shocking, tidal torquing,
and bulge shocking).

\item The clusters in the following sub-sample (viz. NGC~104, NGC~288,
NGC~2298, NGC~5139, NGC~5904,   NGC~6535, and NGC~6809) exhibit  tidal
extensions resulting from a shock, i.e.   tails aligned with the tidal
field gradient.

\item The  clusters in another  sub-sample  (viz.  NGC~1261, NGC~1851,
NGC~1904, NGC~5694,  NGC~5824, NGC~6205,  NGC~7492, Pal~5, and Pal~12)
present  extensions  which are likely  tracing the orbital  path of the
cluster with various degrees of mass loss.

\item NGC~7492 is a striking case because of  its very small extension
and its   high destruction rate driven by   the galaxy as  computed by
GO97.  Its  dynamical ``twin'' for such  an evolution,  namely Pal~12,
exhibits, on the contrary, a large extension tracing its orbital path,
with a possible shock which happened more than 350~Myr.

\item The  velocity diffusion of the stripped  stars is tentatively  
estimated, in one  case  (viz. NGC~6254), to   be similar  to  the 
cluster  velocity dispersion.

\item Thanks to the relatively small distance of NGC~5139 and its high
release of unbound stars during its current disk shocking, we estimate
the  mass loss to be  between 0.6 and  1~\% of the cluster total mass,
taking into account  a possible mass  segregation in the cluster halo.
This  mass loss  rate  is consistent with  our   estimates from N-body
simulations (CLM99).

\item The second break   in the surface  density  slope, apart from
the break at the tidal radius  (cf.   the case  of
NGC~6254) could be an indicator of some recent gravitational shocks,
with the \q{1}{3} indicator displaying a  range of values between -0.9
and -2. The latter is  likely overestimated because of the uncorrected
crowding towards the clusters.
\end{itemize}

The use of  better  quality data,  e.g.  wide-field  CCD observations,
combined  with the  present star-count  method  will  allow in a  near
future to get rid easily of  the observational biases  and to obtain a
better  color selection thanks to   a more accurate photometry.  These
observations will provide  more precise observational estimates of the
mass   loss  rates  for  different regimes   of  galaxy-driven cluster
evolution.  With the help of numerical simulations and accurate proper
motions, it will be  possible to constrain efficiently the  parameters
describing the galactic potential (disk scale-height, surface density,
bulge size).  In  case of a  flat dark  matter  halo (Pfenniger \etal\
1994),  the tidal shock  on the globular  clusters  would be enhanced,
depending on   the surface density  and the  scale-height of this dark
matter halo. Pal~2 is a good candidate to probe  such dark matter halo
flattening, because of its small  distance to the  galactic plane (Z =
-2.2  kpc)  and its relatively large   distance to the galactic center
(21.6 kpc), placing this cluster in a  region where the tidal shocking
by the disk only is expected to be low.

\begin{figure}[hbpt]
\caption{Variations with time of the slope of the mass function fitted
by a power-law (-2.35  for a  Salpeter law).   The simulation  is from
CLM99.}
\label{fig_slope_mf_m1}
\end{figure}

\begin{acknowledgements}
We  acknowledge warmly the  ESO  Schmidt Telescope  operators  O.  and
G. Pizarro, and the efficiency  of the Schmidt astronomers B. Reipurth
and J. Brewer for the conduction of the observations.  We would like to
think the whole MAMA team for its efficient support. We are greatful
to R. Le Poole (Sterrenwacht Leiden) for  facilitating the use of POSS
I   plates.    We   thank E.   Bertin,     L.   Cambresy,  J. Guibert,
C. Loir and M.  Odenkirchen for   very   helpful discussions.   We
acknowledge the  W.E. Harris's catalog  of parameters for the galactic
globular clusters.  O. Gnedin gave  us kindly an electronic version of
his galactic globular cluster destruction rates. And we are very
grateful to the referee, C. Grillmair, for his enlightening
suggestions and comments which helped to improve this paper.
\end{acknowledgements}


\begin{thebibliography}{}
\bibitem{} Abell, G. O., Corwin, G. J., Olowin, P. 1989, ApJS, 70, 1
\bibitem{} Alcaino, G., Liller, W. 1987, AJ, 94, 1585
\bibitem{} Aguilar, L. A., Hut, P., Ostriker, J.P. 1988, ApJ, 335,720
\bibitem{} Anderson, J., 1998, PhD thesis, Berkeley University
\bibitem{} Anderson, J., King, I. R. 1996, ASP Conf. Ser. 92: Formation of 
the Galactic Halo...Inside and Out, 257 
\bibitem{} Antonov V.A. 1962, Vest. leningr. gos. Univ., 7, 135
\bibitem{} Armandroff, T. E., Da Costa, G. S. 1991, AJ, 101, 1329
\bibitem{} Bahcall, J. N., Soneira, R. M.,  Schmidt, M. 1983, ApJ, 265, 730
\bibitem{} Berger, J., Cordoni, J. P., Fringant, A. M., Guibert, J., Moreau, O., 
           Reboul, H., Vanderriest, C. 1991, A\&AS, 87, 389 
\bibitem{} Bergond, G., Leon, S., Guibert, J. 2000, in prep.
\bibitem{} Bertin, E., Arnouts, S. 1996, A\&AS, 117, 393
\bibitem{} Binney, J., Tremaine, S., 1987, in ``Galactic Dynamics'', Princeton Series
\bibitem{} Bijaoui, A. 1991, ``The Wavelet Transform'', Data Analysis Workshop, 3
\bibitem{} Caldwell, J. A. R., Ostriker, J. P. 1981, ApJ, 251, 61 
\bibitem{} Cambr\'ezy, L. 1999, A\&A, 345, 965
\bibitem{} Cannon, R.D., Stobbie, R.S., 1973, MNRAS, 162, 207
\bibitem{} Chernoff, D.F., Kochanek, C.S., Shapiro, S.L. 1986, ApJ, 309, 183
\bibitem{} Chereul, E., Cr\'ez\'e, M., Bienaym\'e, O. 1999, A\&A, 135, 5
\bibitem{} Combes, F., Leon, S., Meylan., S., 1999, A\&A, 352, 149 (CLM99)
\bibitem{} Dinescu, D. I., Girard, T. M., Van Altena, W. F., Mendez, R. A., 
           Lopez, C. E. 1997, AJ, 114, 1014
\bibitem{} Djorgovski S.G., Meylan G. 1994, 108, 1292
\bibitem{} Geisler, D., Piatti, A. E., Claria, J. J., Minniti, D. 1995, AJ, 109, 605
\bibitem{} Gill, A.G., Henriksen, R.N. 1990, ApJ, 365, 27
\bibitem{} Gillett, F. C., Neugebauer, G., Emerson, J. P., Rice, W. L. 1986, ApJ, 300, 722 
\bibitem{} Gnedin, O.Y., Ostriker, J.P. 1997, ApJ, 474, 223  (GO97)
\bibitem{} Goodman, J., Hut, P. 1989, Nature, 339, 40 
\bibitem{} Grillmair, C.J. 1992, PhD thesis, Australian National University
\bibitem{} Grillmair, C. J., Ajhar, E. A., Faber, S. M., Baum, W. A., 
           Holtzman, J. A., Lauer, T. R., 
           Lynds, C. R., O'Neil, E. J., J.. 1996, AJ, 111, 2293 
\bibitem{} Grillmair, C. J., Freeman, K. C., Irwin, M.,  Quinn, P. J. 1995, AJ, 109, 2553 
\bibitem{} Grossmann, A., Morlet, J 1985, in Streit L. (ed) Mathematics+Physics, Lecture on 
recent results. World Scientific, Singapore
\bibitem{} Gunn, J.E., Griffin, R.F. 1979, AJ, 84, 752
\bibitem{} Harris, W.E. 1996, AJ, 112, 1487 
\bibitem{} Hesser, J. E., Shawl, S. J., Meyer, J. E. 1986, PASP, 98, 403
\bibitem{} Hurley, D. J. C., Richer, H. B., Fahlman, G. G. 1989, AJ, 98, 2124
\bibitem{} Johnston, K. V., Sigurdsson, S., Hernquist, L. 1998, astro-ph/9805291
\bibitem{} Kundic, T. Ostriker, J. P. 1995, ApJ, 438, 702 
\bibitem{} Lehmann, I., Scholz, R.-D. 1997, A\&A, 320, 776
\bibitem{} Leon, S., Combes, F., Leeuwin F. 2000, in prep.
\bibitem{} Leon, S., Bergond, G., Vallenari, A., 1999, A\&A, 344, 450
\bibitem{} Longaretti, P.-Y., Lagoute, C. 1996, 308, 453
\bibitem{} Lynden-Bell D., Wood, R., 1968, MNRAS, 138, 495
\bibitem{} Merritt, D., Meylan, G., Mayor, M., 1997, AJ, 114, 1074-1086
\bibitem{} Meylan, G., 1987, A\&A, 184, 144-154
\bibitem{} Meylan, G., 1989, A\&A, 214, 106-112
\bibitem{} Meylan, G., Mayor, M. 1986, A\&A, 166, 122 
\bibitem{} Meylan, G., Mayor, M., Duqueynnoy, A., Dubath, P. 1995, A\&A, 303, 761
\bibitem{} Meylan, G., Mayor, M. 1991, A\&A, 250, 113
\bibitem{} Meylan, G., Heggie, D.C., 1997, A\&AR, 8, 1
\bibitem{} Moore, B., 1996, ApJ, 461, L13
\bibitem{} Mould, J. R., Watson, A. 
M., Gallagher, J. S., I., Ballester, G. E., Burrows, C. J., Casertano, S., 
Clarke, J. T., Crisp, D., Griffiths, R. E., Hester, J. J., Hoessel, J. G., 
Holtzman, J. A., Scowen, P. A., Stapelfeldt, K. R., Trauger, J. T., Westphal, J. A. 1996, PASP, 108, 682 
\bibitem{} Murali, C.  Weinberg, M. D. 1997, MNRAS, 291, 717 
\bibitem{} Murtagh, F., Starck, J. -L., Bijaoui, A. 1995, A\&AS, 112, 179 
\bibitem{} Odenkirchen, M., Brosche, P., Geffert, M., Tucholke, H-J.,
           1997, New Astronomy, 2, 477 
\bibitem{} Odenkirchen, M. 1998, private communication
\bibitem{} Oh, K.S., Lin, D.N.C. 1992, ApJ, 386, 519
\bibitem{} Ostriker, J.P. \& Caldwell, J. A. 1983, in Kinematics,
Dynamics and Structure of the Milky Way, ed. W. L. H. Shuter (Dordrecht: Reidel), 24
\bibitem{} Pfenniger, D., Combes, F., Martinet, L. 1994, AA, 285, 79 
\bibitem{} Portegies Zwart, S. F., Hut, P., Verbunt, F. 1997, A\&A, 328, 130 
\bibitem{} Vesperini, E., Heggie, D. C. 1997, MNRAS, 289, 898 
\bibitem{} Pryor, C., Meylan, G. 1993, in ``Structure and dynamics of globular clusters''
           Djorgovski \& Meylan (eds), A.S.P. Conf. Ser. Vol 50
\bibitem{} Saviane, I., Piotto, G., Fagotto, F., Zaggia, S., Cappacioli, M., 
           Aparicio, A. 1998, A\&A, 333, 479
\bibitem{} Scholz, R. -D., Irwin, M., Odenkirchen, M., Meusinger, H. 1998, A\&A, 333, 531 
\bibitem{} Slezak, E., Durret, F., Gerbal, D. 1994, AJ, 108, 1996 
\bibitem{} Schweitzer, A. E., Cudworth, K. M., Majewski, S. R. 1993, ``The Globular 
           Cluster-Galaxy Connection. Globular Clusters Within the Context of Their 
           Parent Galaxies'', ASP, p. 113
\bibitem{} Spitzer, L.J. 1958, ApJ, 127, 17
\bibitem{} Spitzer, L.J., Hart, M.H. 1971, ApJ, 166, 483
\bibitem{} Spitzer, L.J., Chevalier R.A. 1973, 183, 565
\bibitem{} Starck, J.-L, Siebenmorgen, R., Gredel, R. 1997, ApJ, 482, 1011 
\bibitem{} Trager S.C., King I.R., Djorgovski S. 1995, AJ, 109, 218
\bibitem{} Vesperini, E. Heggie, D. C. 1997, MNRAS, 289, 898 
\bibitem{} Webbink, R. F. 1981, ApJS, 45, 259 
\bibitem{} Weinberg, M.D. 1994, AJ, 108, 1414
\bibitem{} Zaggia, S.R., Piotto, G., Capaccioli, M. 1997, A\&A, 324, 1004
\bibitem{} Zoccali, M., Piotto, G., Zaggia, S. R., Capaccioli, M. 1998, A\&A, 331, 541 
\end{thebibliography}
\end{document}